\documentclass[conference]{IEEEtran}
\IEEEoverridecommandlockouts
\usepackage{cite}
\usepackage{amsmath,amssymb,amsfonts}
\usepackage{algorithmic}
\usepackage{graphicx}
\usepackage{textcomp}
\usepackage{url} 
\usepackage{stmaryrd}
\usepackage{cleveref}
\usepackage{braket}
\usepackage{bm}
\usepackage{adjustbox}
\usepackage{subcaption}
\usepackage{multirow}
\usepackage[table]{xcolor}
\usepackage{caption} 
\usepackage{algorithm}
\usepackage{comment}

\newcommand{\update}[1]{{{\color{black}#1}}}
\newcommand{\RNum}[1]{\uppercase\expandafter{\romannumeral #1\relax}}
\def\BibTeX{{\rm B\kern-.05em{\sc i\kern-.025em b}\kern-.08em
    T\kern-.1667em\lower.7ex\hbox{E}\kern-.125emX}}

\begin{document}

\title{Iceberg Beyond the Tip: Co-Compilation of Quantum Error Detection Codes and Quantum Algorithms}

\author{                                  
      \IEEEauthorblockN{Yuwei Jin\IEEEauthorrefmark{1}\IEEEauthorrefmark{4},                                                             
                        Zichang He\IEEEauthorrefmark{1}\IEEEauthorrefmark{4},                                       
                        Tianyi Hao\IEEEauthorrefmark{1}\IEEEauthorrefmark{3},
                        Sivaprasad Omanakuttan\IEEEauthorrefmark{1}}
      \IEEEauthorblockN{MinZhao Liu\IEEEauthorrefmark{1},
                        David Amaro\IEEEauthorrefmark{2},
                        Swamit Tannu\IEEEauthorrefmark{3},
                        Ruslan Shaydulin\IEEEauthorrefmark{1}}
      \IEEEauthorblockN{Marco Pistoia\IEEEauthorrefmark{1}}
      \IEEEauthorblockA{\IEEEauthorrefmark{1}Global Technology Applied Research, JPMorganChase, New York, NY 10001, USA}
      \IEEEauthorblockA{\IEEEauthorrefmark{2}Quantinuum, Partnership House, Carlisle Place, London SW1P 1BX, UK}
      \IEEEauthorblockA{\IEEEauthorrefmark{3}Department of Computer Sciences, University of Wisconsin-Madison, Madison, WI 53706, USA}
      \thanks{\IEEEauthorrefmark{4}These authors contributed equally to this work.}
  }

\maketitle

\begin{abstract}
Recent advances in quantum hardware have enabled logical-qubit demonstrations and system-scale experiments that turn quantum processors into practical platforms for algorithmic benchmarking. 
A central challenge in leveraging these devices is mitigating the impact of noise, which can obscure algorithmic observables. 
Quantum error detection (QED) codes, such as the $\llbracket k\!+\!2,k,2\rrbracket$ Iceberg code, address this by identifying and discarding erroneous runs. 
However, conventional compilation flows maintain fixed gadget structures and are unable to exploit the flexibility in QED gadget encoding and ordering. This limitation restricts parallelism and increases qubit idle exposure, which is particularly critical for trapped-ion devices.
We present a hardware-aware co-compilation framework that jointly optimizes Iceberg QED gadgets and algorithmic circuits to reduce two-qubit depth and idling while preserving gadget-level fault tolerance. 
Evaluating on Quantinuum H2-1 trapped-ion hardware, we show that for MaxCut QAOA our co-compilation reduces two-qubit depth by up to $55\%$ relative to fixed-gadget baselines, increases post-selection rate from $4\%$ to $33\%$, and improves success probability from $44\%$ to $65\%$ on the same $k=22$ instance with $330$ algorithmic and $744$ physical two-qubit gates. 
Furthermore, we demonstrate beyond-break-even performance for up to $34$ algorithmic qubits, employing $510$ algorithmic two-qubit gates and $1140$ physical two-qubit gates.
We additionally show the effectiveness of our co-compilation on other applications, including instantaneous quantum polynomial (IQP) circuits and quantum Fourier transform (QFT) circuits.
These results highlight the potential of co-compilation to advance fault-tolerant quantum computation on near-term hardware.
\end{abstract}

\section{Introduction}

Today's quantum computers are capable of performing computational tasks that surpass the reach of even the most powerful classical supercomputers~\cite{Morvan2024,qntm_rcs,Gao2025,Liu2025}.
This capability creates exciting opportunities to use near-term quantum devices as platforms for benchmarking and designing algorithms, accelerating progress toward commercial quantum advantage. Realizing this potential, however, depends on our ability to extract accurate estimates of algorithmic observables—such as expected performance or success probability—from inherently noisy quantum experiments.

To address the challenge of noise, a variety of error suppression techniques have been developed. These methods, including probabilistic error cancellation~\cite{vandenBerg2023,Kim2023}, typically involve repeated execution of modified circuits and have demonstrated practical success, even for circuits with more than 100 qubits. Among these approaches, QED stands out for its ability to improve circuit fidelity across diverse hardware platforms~\cite{2409.04628,2404.02280,2412.15165,Gupta2024}, enabling the execution of complex algorithmic circuits~\cite{Self2024,Bluvstein2023,he2024performance,wang2024fault,reichardt2024logical,pokharel2024better,ginsberg2025quantum}. QED is not only complementary to other signal extraction techniques, such as probabilistic error cancellation and zero-noise extrapolation~\cite{2501.09079}, but also serves as a foundational element in scalable fault-tolerant architectures, including magic state preparation~\cite{Gupta2024,chamberland2019fault} and distillation~\cite{Bravyi2012}.

Protecting an algorithmic circuit with an error-detecting code requires compiling it into a larger circuit that integrates the encoded algorithm with fault-tolerant gadgets, such as state preparation and syndrome measurement. While this expansion introduces additional opportunities for errors, error detection gadgets mitigate these risks by identifying and discarding faulty runs. The central objective is to achieve ``beyond-break-even'' (better-than-unencoded) performance, where the encoded circuit delivers superior results compared to its unencoded counterpart.

Standard compilation practices typically encode a pre-optimized algorithmic circuit using fixed QED gadgets~\cite{autobraid,qft_qec}, but this approach overlooks the inherent flexibility in both encoding and gadget construction. Moreover, conventional circuit optimization techniques cannot be directly applied to encoded circuits, as the fault-tolerance property of the gadgets must be preserved~\cite{qaoaMappingAstar,linlin:2qan,gushu:paulihedral,liu:quclear,jin:tetris,phasePolySynthesis,UnitarySynthesis,qfastUnitarySynthesis}. %
These limitations are particularly pronounced in trapped-ion quantum computers, where memory errors, which arise from qubits idling and decohering, pose a major obstacle to high-fidelity computation~\cite{Moses2023,Reichardt2024}. 
As circuit sizes and depths increase, memory errors become increasingly significant, and the substantial overhead introduced by fault-tolerant encoding further exacerbates the challenge. 

\begin{figure}[t]
    \centering
    \includegraphics[width=0.85\linewidth]{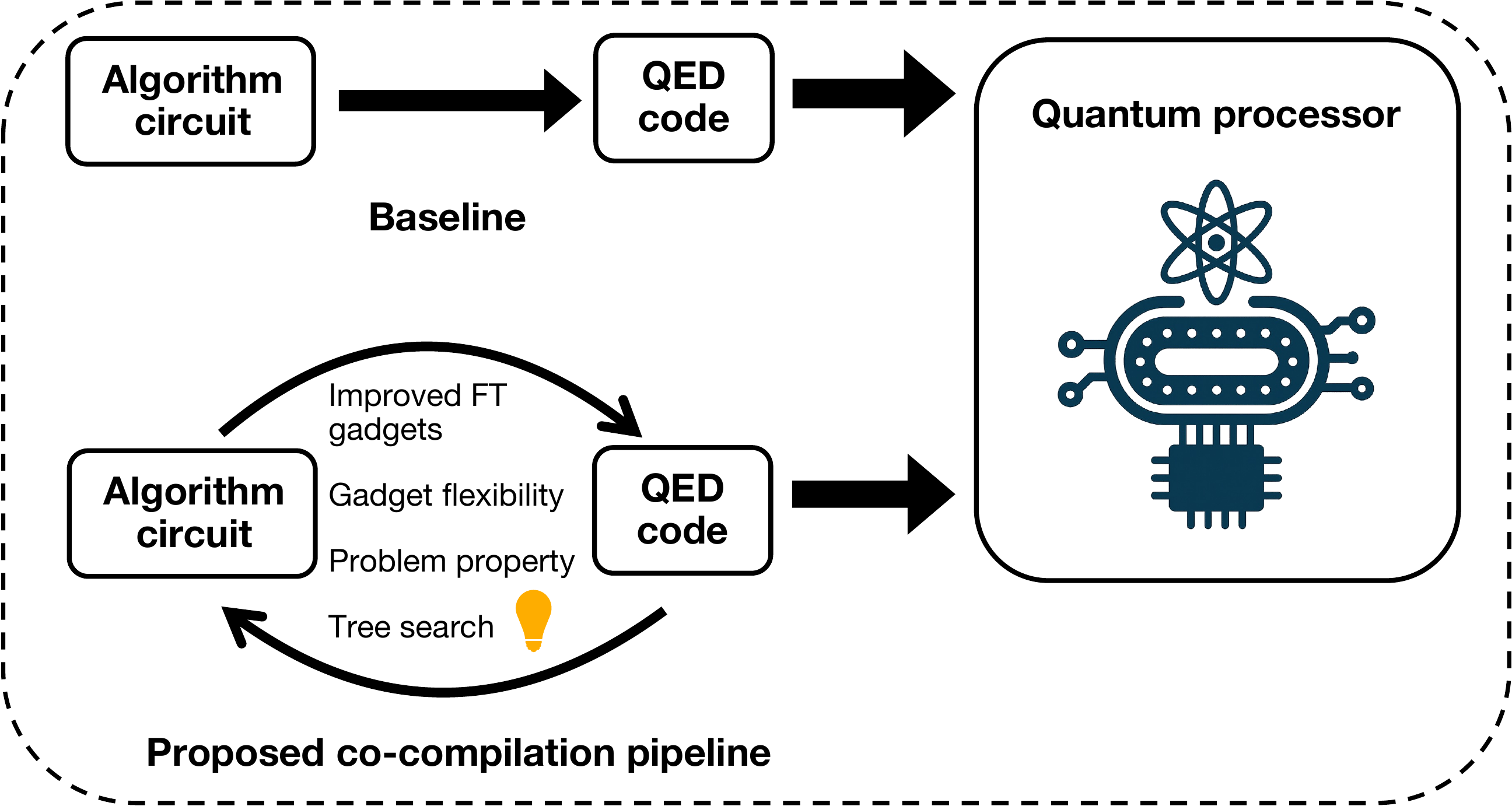}
    \caption{
    Overview of the co-compilation pipeline for quantum algorithms and quantum error detection. 
    The figure illustrates how algorithmic circuits are integrated with flexible fault-tolerant gadgets from the error detection code.
    By jointly optimizing the structure and ordering of algorithmic gates and error detection gadgets, the pipeline enables more efficient circuit synthesis for near-term quantum hardware.} %
    \label{fig:first_panel}
\end{figure}
This work introduces a co-compilation pipeline that jointly optimizes both the algorithmic circuit and the error detection code gadgets, as illustrated in \Cref{fig:first_panel}. By leveraging the flexibility in both components, this approach enables more efficient circuit synthesis, advancing beyond-break-even performance on near-term quantum hardware.

As an example, we consider the compilation of QAOA circuits protected by the $\llbracket k+2, k, 2 \rrbracket$ Iceberg error-detecting code, though the methods developed here are broadly applicable to other algorithmic circuits. 
QAOA is a promising quantum algorithm, with evidence of exponential speedups for certain problem classes~\cite{qaoa_near_symmetric_optimization_problems,2503.12789}, and its moderate resource requirements make it well-suited for near-term and early fault-tolerant hardware~\cite{Shaydulin2023npgeq,Pelofske2023,Pelofske2024,Tasseff2024,he2023alignment,he2024performance,omanakuttan2025threshold}. 
Beyond QAOA, we also demonstrate the co-compilation pipeline on instantaneous quantum polynomial (IQP) circuits and Quantum Fourier transform (QFT) circuits.
The Iceberg code is particularly well-matched to QCCD trapped-ion quantum computers, thanks to its compatibility with all-to-all connectivity and native gate support~\cite{Self2024}, while it has also been applied to other hardware platforms~\cite{pattison2024fast,zhong2025combining}. Although the Iceberg code has enabled beyond-break-even results in several applications~\cite{nishi2025encoded,yamamoto2024demonstrating,van2024end,ginsberg2025quantum,he2024performance}, achieving beyond-break-even performance for algorithmic circuits with more than 20 qubits remains a challenge, underscoring the need for further optimization.

Our first contribution is the identification of optimization opportunities for QED-encoded circuits and the introduction of a co-compilation pipeline.  %
We exploit the underlying flexibility of QED gadgets to enable co-optimization with algorithmic circuits while maintaining fault-tolerance. 
For our QAOA example, we
further leverage a new logical gate implementation enabled by the $\mathbb{Z}_2$ symmetry present in many problems, such as MaxCut, where solutions remain invariant under global bit-flip.

Our second contribution is the development of a tree search method to automate the co-optimization pipeline. It employs graph representations to capture all executable gate options arising from the flexibility in the algorithm and the Iceberg gadgets. Systematic exploration of this graph identifies gate combinations and their impact on circuit depth, guided by a search algorithm. %

Our third contribution is the experimental demonstration of QAOA on hardware, extending beyond previous state-of-the-art results. Beyond-break-even performance is observed for circuits with up to 34 algorithmic (logical) qubits and $510$ algorithmic two-qubit gates. At this scale, classical simulation of the circuit becomes challenging. Comparison with prior large-scale QAOA experiments on MaxCut reveals substantial improvements across all tested circuits. Notably, the previous largest break-even point of 20 algorithmic qubits and $300$ algorithmic two-qubit gates~\cite{he2024performance} is surpassed. The enhanced hardware performance is attributed to optimizations that reduce circuit depth by up to $55\%$ relative to the baseline Iceberg code implementation~\cite{he2024performance}.

In addition, we demonstrate an application of using the Iceberg code to benchmark QAOA energy populations on hardware. The Iceberg code shows promise in capturing QAOA energies under noisy conditions. Inspired by the long-tailed hardware results, we illustrate that a simple post-processing strategy can effectively bring the Iceberg energy populations closer to the noiseless distribution.

\section{Background}

\subsection{Quantum Error Correction and Detection}

Quantum computers are fundamentally limited by the fragility of quantum information. Qubits are highly susceptible to errors. Strategies that protect quantum information from such errors are essential.
Quantum Error Correction (QEC) involves encoding logical qubits into multiple physical qubits to correct errors.
However, current quantum hardware faces challenges in implementing QEC \cite{gidney2025factor2048bitrsa,omanakuttan2025threshold,zhou2025resource}.

Quantum Error Detection (QED) is a related but less resource-intensive approach, where the code is designed to detect, but not necessarily correct, certain classes of errors. When an error is detected, the corresponding computational run is discarded. While this post-selection approach does not recover lost information, it enables the extraction of high-fidelity results from repeated experiments, provided the error rate and discard overhead are manageable. QED is particularly attractive for near-term devices, where the overhead of full error correction is prohibitive. For a more detailed review of QEC and QED, see \cite{gottesman2009introductionquantumerrorcorrection,lidar2013quantum}.

\subsection{The Iceberg Code: Structure and Properties}
The Iceberg code~\cite{Steane1996, Gottesman1998, Self2024} is a quantum error detection code specifically designed to efficiently detect single-qubit errors while minimizing resource overhead. It is denoted as a $\llbracket k+2, k, 2 \rrbracket$ code, meaning it encodes $k$ logical qubits into $k+2$ physical qubits and has distance 2, i.e., it can detect any single-qubit error but not necessarily correct it.

\begin{figure}[t]
    \centering
    \includegraphics[width=0.9\linewidth]{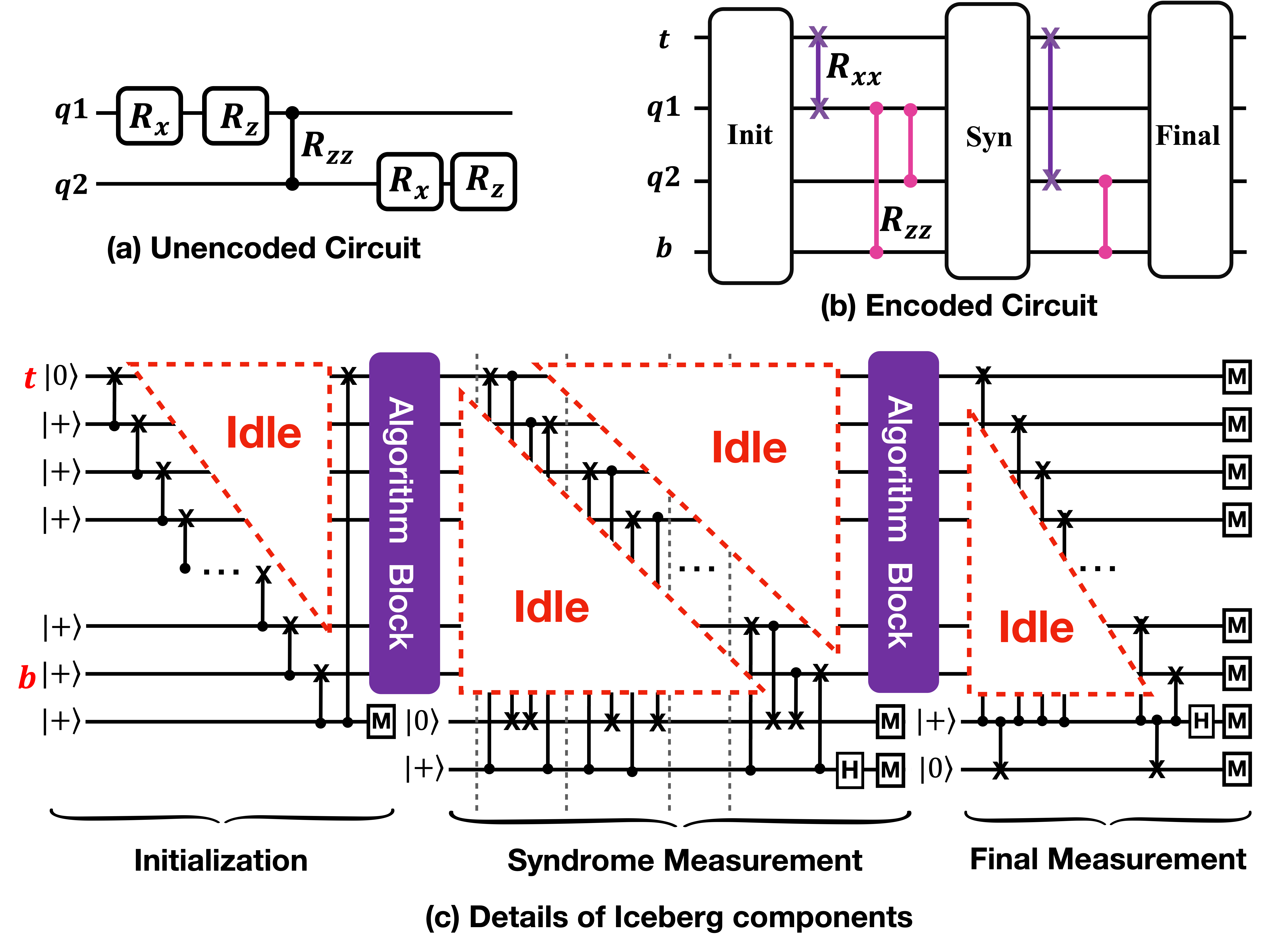}
    \caption{ Examples of Iceberg code circuits. 
    (a) An unencoded circuit with single- and two-qubit gates.
    (b) The Iceberg-encoded circuit: single-qubit rotation gates are replaced by two-qubit rotation gates, while the two-qubit gate remains unchanged.
    (c) Detailed schematic of Iceberg code gadgets. Extra ancilla qubits are used for syndrome detection and are shown below the bottom $b$ qubit. In the syndrome gadget, each qubit ($t$, $b$, $1 \ldots k$) is connected to two separate ancilla qubits. The circuit encoded with the Iceberg code, building on prior art~\cite{he2024performance,Self2024}, presents numerous opportunities for further optimization, which we explore in this work. }
    \label{fig:motivate_encoded_circuit}
\end{figure}

The code is defined by two stabilizer operators:
\begin{align*}
    S_z = Z_t Z_b \prod_{i=0}^k Z_i,  \quad  S_x = X_t X_b \prod_{i=1}^k X_i,
\end{align*}
where $Z_i$ and $X_i$ are Pauli operators acting on the $i$-th qubit, and $t$ and $b$ denote the top and bottom qubits (see \Cref{fig:motivate_encoded_circuit} for detailed description). 
The code space is the simultaneous $+1$ eigenspace of both $S_z$ and $S_x$. 
Any single-qubit Pauli error will anti-commute with at least one of the stabilizers, causing a detectable change in the measurement outcome.

A key feature of the Iceberg code is its support for efficient, partially fault-tolerant logical gate implementations. Logical Pauli operators are mapped to collective operations on the physical qubits, and arbitrary-angle logical Pauli rotations $\exp(-i\theta \overline{P})$ for $\overline{P} \in \{\overline{X}, \overline{Z}, \overline{XX}, \overline{YY}, \overline{ZZ}\}$ can be implemented using only a single two-qubit physical gate\cite{Self2024}. 
For example, a logical $\overline{X}$ rotation on logical qubit $i$ is implemented as a physical $X_t X_i$ rotation ($R_{xx}$),  and a logical $\overline{Z}$ rotation on logical qubit $i$ is implemented as a physical $Z_b Z_i$ rotation ($R_{zz}$). The two-qubit Pauli rotation gates remain the same. Because a $\overline{Z_iZ_j}$ = $Z_b Z_iZ_bZ_j = Z_iZ_j$, the same for logical $\overline{X_iX_j}$ rotation gate. A concrete example is shown in \Cref{fig:motivate_encoded_circuit}. 

The Iceberg code also has efficient gadgets for fault-tolerant initialization, syndrome measurement, and final measurement~\cite{Self2024}. 
It is important to note that, as a distance-2 code, the Iceberg code is only partially fault-tolerant: while it detects all single-qubit errors, certain two-qubit errors (specifically, $XX$, $YY$, or $ZZ$ on pairs of qubits) may remain undetected. Nevertheless, for many near-term applications, the ability to efficiently detect and discard runs affected by single-qubit errors provides a significant improvement in overall fidelity.

The efficiency of the Iceberg code is quantified by the \emph{post-selection rate}, defined as the fraction of experimental runs that pass all stabilizer checks and are retained for analysis. 
This rate depends on the physical error rate, the circuit depth, and the number of qubits, and serves as a key metric for the practical viability of QED-based approaches.

\update{Beyond the Iceberg code, the co-compilation and gadget-resynthesis techniques developed here are broadly applicable to other quantum error-detection codes that employ high-weight stabilizers. In particular, the same approach extends to H codes with parameters $\llbracket k+4, k, 2\rrbracket$, which, like the Iceberg family, implement high-weight stabilizer measurements~\cite{jones2013multilevel}. Likewise, these compilation techniques apply to concatenation-based codes with high-weight stabilizers, which have attracted significant attention for fault-tolerant quantum computation~\cite{yoshida2025concatenate,goto2024high}.}

\subsection{Quantum Approximate Optimization Algorithm}

To concretely study the interplay between error detection, memory errors, and circuit structure, we focus on the QAOA~\cite{Hogg2000,Hogg2000search,farhi_qaoa}, a leading candidate for demonstrating quantum advantage in combinatorial optimization.
QAOA prepares a parameterized quantum state by alternating between mixing and phase-separating operators, with the goal of finding a bitstring that minimizes a given objective function.

Formally, for a cost function $f$ defined on the Boolean hypercube and encoded as a Hamiltonian $C$ such that $C\ket{x} = f(x)\ket{x}$ for all $x \in \{0,1\}^n$, the QAOA state is given by
\begin{align}
    \ket{\psi_p\left(\bm\beta, \bm\gamma\right)} = \prod_{t = 1}^p e^{-i\beta_t\sum_{1 \leq j \leq n}X_j}e^{-i\gamma_tC}\ket{+}^{\otimes k},
    \label{eq:def_qaoa_state}
\end{align}
where $X_j$ is the Pauli $X$ operator on qubit $j$, $p$ is the number of alternating layers (the QAOA depth), and $\bm\beta, \bm\gamma$ are sets of variational parameters. In this work, we focus on the MaxCut problem, where the cost Hamiltonian is
\begin{align}
    C = \sum_{(i,j)\in E} Z_i Z_j,
\end{align}
with $G=(V,E)$ the input graph. For many problem classes, effective parameter choices are known~\cite{qaoa_labs,qaoa_weighted_maxcut,qaoa_ksat,qaoa_sk,Wurtz2021,2503.12789,ICCAD_qaoapara,hao2024end,he2025non}, and for MaxCut on regular graphs, we use fixed parameters from~\cite{Wurtz2021}.  In this work, we use the MaxCut problem as a representative application for QAOA, encoding the problem Hamiltonian into the quantum circuit and studying the performance of Iceberg-encoded QAOA circuits under realistic noise and error models.

An example of QAOA circuit with $p = 1$ is shown in \Cref{fig:gadgetOpt_qaoa}(a)-(c). Each edge from the input graph is an $R_{zz}$ rotation gate in the circuit and the gate order is flexible. This flexibility allows us to synthesize the logical circuit with less logical circuit depth. \Cref{fig:gadgetOpt_qaoa}(b) and (c) are two valid QAOA circuits for the input problem graph in \Cref{fig:gadgetOpt_qaoa}(a). The encoded QAOA cost Hamiltonian and mixer Hamiltonian are shown in \Cref{fig:gadgetOpt_qaoa}(d) and \Cref{fig:gadgetOpt}(a). The encoded cost Hamiltonian remains the same; however, a set of single-qubit $R_x$ gates in QAOA mixer are encoded as $R_{xx}$ gates. They share the same top qubit, resulting in the circuit depth increasing by a factor of $k$.

\section{Motivation and Insights}
\label{sec:insights}

To achieve ``beyond-break-even'' performance, it is essential to mitigate overhead, with a particular focus on memory errors within the encoded circuit. In this section, we will first highlight the importance of protecting against memory errors and introduce three major optimization opportunities.

\subsection{Memory Error Matters}

While error detection and correction codes are essential for mitigating noise, their practical effectiveness is significantly shaped by the structure and depth of the quantum circuit.
Large or non-optimized circuits can experience substantial \emph{memory errors} from idle qubits exposed to decoherence \cite{gidney2025factor2048bitrsa}. In encoded circuits, qubits not constantly participating in gate operations become susceptible to such errors. The likelihood of memory errors escalates with the number of idle qubits and extended idling durations. 
Consequently, higher-depth circuits with prominent qubit idling often amass undetected, uncorrected errors, compromising computational fidelity. 
As illustrated in \Cref{fig:motivate_encoded_circuit}(c) for Iceberg code, the encoded circuits may exhibit considerable idle regions. For quantum algorithms with numerous single-qubit gates, the logical implementation necessitates frequent use of either the top or bottom qubit, substantially increasing circuit depth.

To reveal the impact of memory errors, we run Iceberg-encoded QAOA using the baseline strategy~\cite{he2024performance} on the Quantinuum H2-1 emulator, where circuits are optimized by tket~\cite{sivarajah2020t} with barriers inserted for the fault-tolerant gadgets. In addition to the post-selection rate, we employ two metrics. The logical fidelity~\cite{he2024performance} measures the gap between noiseless and noisy performance of the QAOA executions. The space-time area is defined as $(k+2) \times \text{2Q depth}$, where 2Q depth is the depth of two-qubit gates. Higher space-time area leads to higher probability of getting memory errors. To control the impact of gate errors, we choose the number of algorithmic qubits $k$ and depth $p$ in the QAOA circuit such that the instances have similar numbers of physical two-qubit gates in the encoded QAOA circuit. Despite similar contributions of gate errors, as the space-time area increases, the logical fidelity and the post-selection rate of the circuit decrease quickly, as shown in \Cref{tab:motivate_depth}. Similar observations have also been made in~\cite{he2024performance, Ginsberg2025}.
This indicates that in large-scale circuits, the impact of memory errors on the circuit performance is non-negligible, which motivates us to develop better compilation techniques for the QED-encoded circuit to reduce qubit idling and circuit depth. 
\begin{table}[!t]
\centering
\begin{adjustbox}{width=0.9\linewidth}
\begin{tabular}{c|ccccc}
\hline
$(k,p)$  & $(18,10)$ & $(20,9)$ & $(22,8)$ & $(24,7)$ \\
\hline
\# 2Q Gates   & 615 & 631 & 637 & 633 \\
2Q depth  & 375 & 389 & 397 & 403  \\
Space-time area & 7500 & 8558 & 9528 & 10478 \\
\hline
Post-selection rate (\%)   & 13.9 $\pm 0.6$ & 10.9 $\pm 0.6$ & 8.3 $\pm 0.5$ & 6.6 $\pm 0.4$ \\
\hline
\multirow{2}{*}{Logical fidelity}  & $0.945$  &  $0.899$ & $0.889$ & $0.801$  \\
& \small{$\pm 0.011$}  &  \small{$\pm 0.018$} & \small{$\pm 0.019$} & \small{$\pm 0.028$}  \\
\hline
\end{tabular}
\end{adjustbox}
\caption{The performance of QAOA encoded using the Iceberg code into circuits with a similar number of physical two-qubit gates highlights the importance of space-time area for algorithmic performance. These circuits were emulated using the Quantinuum H2-1 emulator with a total of 3000 shots. The error bars represent the standard error arising from the limited number of post-selected samples.}
\label{tab:motivate_depth}
\end{table}

\subsection{\update{New Set of Gadgets Design}}
\update{
In this work, we propose optimized versions of the fault-tolerant gadgets (\Cref{fig:newGadgets}) of the Iceberg code that consume less circuit depth and fewer two-qubit gates. The FT property of the new gadgets can be easily verified by enumerated error propagation. We report the comparisons in \Cref{tab:gadget_depth}.

The new fault-tolerant initialization gadget in \Cref{fig:newGadgets}(a) (also proposed in~\cite{Goto2024} during the development of this work) halves the depth by using a two-branch GHZ construction and measuring the parity of the two end qubits of each branch.
The depth can be further reduced by using more branches, but preserving the fault-tolerance would require measuring more qubit parities, consuming additional ancillary qubits and two-qubit gates. 
The fault-tolerant initializion gadget for preparing the initial logical state $\ket{\overline{+}}^{\otimes k}$ of QAOA differs from the gadget in~\cite{Self2024}, that prepares $\ket{\overline{0}}^{\otimes k}$, by a transversal physical Hadamard gate that can be pulled all the way to the qubit initialization. 
The new syndrome measurement in \Cref{fig:newGadgets}(b) improves the depth of the original syndrome measurement from $k+6$ to just $k+2$ by carefully swapping the order of the CNOTs. 
Importantly, this new gadget is only valid for a number $n=k+2$ of physical qubits that is a multiple of 4, as otherwise it entangles the ancillas. 
The new final measurement in \Cref{fig:newGadgets}(c) reduces by 1 the number of ancillas and CNOTs.
\begin{figure}[t]
    \centering
    \includegraphics[width=0.5\textwidth]{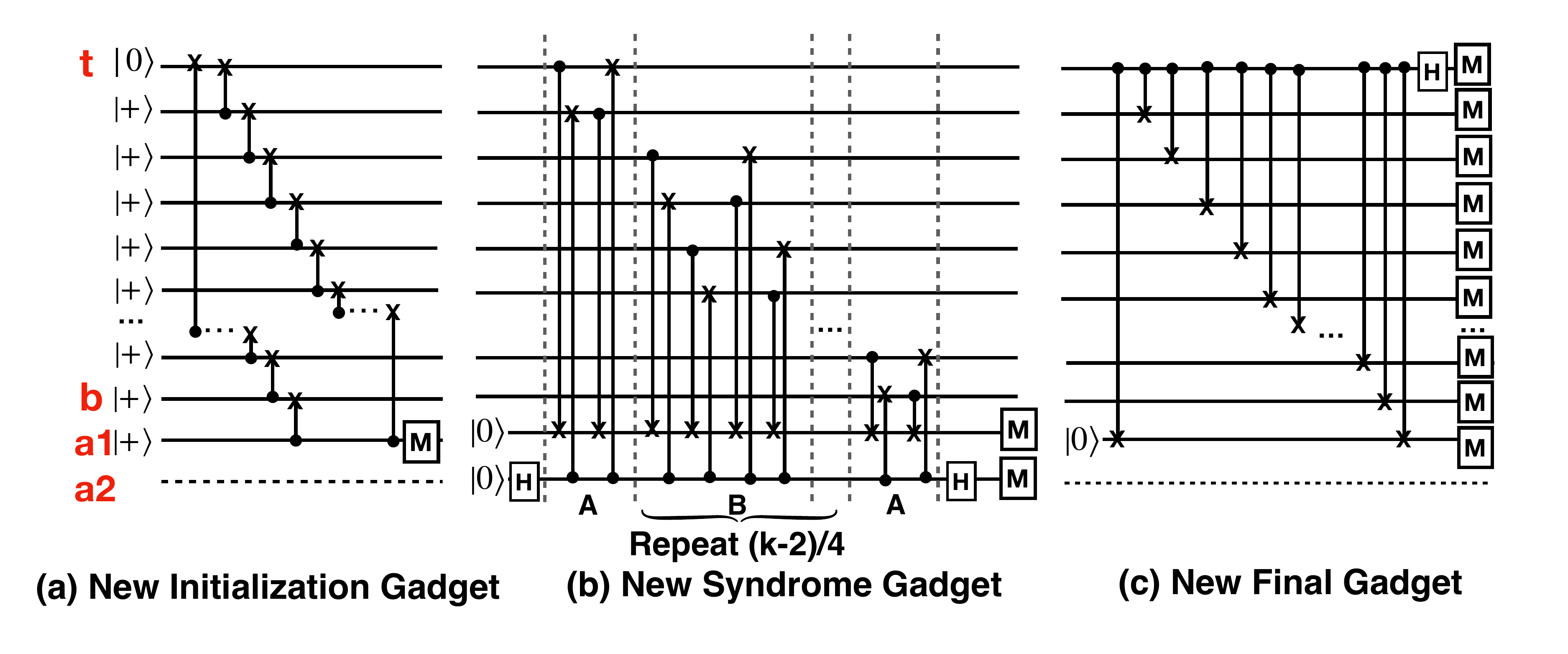}
    \caption{
    \update{New set of fault-tolerant gadgets. (a) New initialization gadget for preparing the logical $\ket{\overline{+}}^{\otimes k}$. (b) New syndrome measurement gadget with higher parallelism. (c) New final measurement gadget with one physical qubit less. }
    }
    \label{fig:newGadgets}
\end{figure}

\begin{table}[]
\centering
\begin{adjustbox}{width=\linewidth}
\begin{tabular}{c|ccccc}
\hline
&Gadgets  & Initialization & Syndrome gadget & Final gadget \\
\hline
\multirow{2}{*}{2q\_depth} &Previous~\cite{Self2024} & $k+3$ & $k+6$ & $k+4$ \\
&Proposed & $k/2+3$ & $k+2$ & $k+3$ \\
\hline
\multirow{2}{*}{2q\_gate} &Previous~\cite{Self2024} & $k+3$ & $2k+4$ & $k+4$ \\
&Proposed & $k+3$ & $2k+4$ & $k+3$ \\
\hline
\end{tabular}
\end{adjustbox}
\caption{\update{Comparison on depth of different FT gadgets for the number $k$ of algorithmic qubits. Note that the proposed gadgets here have not revealed the depth reduction of co-compilation of algorithmic circuit yet. }
}
\label{tab:gadget_depth}
\end{table}

}

\subsection{Gadget Resynthesis}

\begin{figure}[t]
    \centering
    \includegraphics[width=0.9\linewidth]{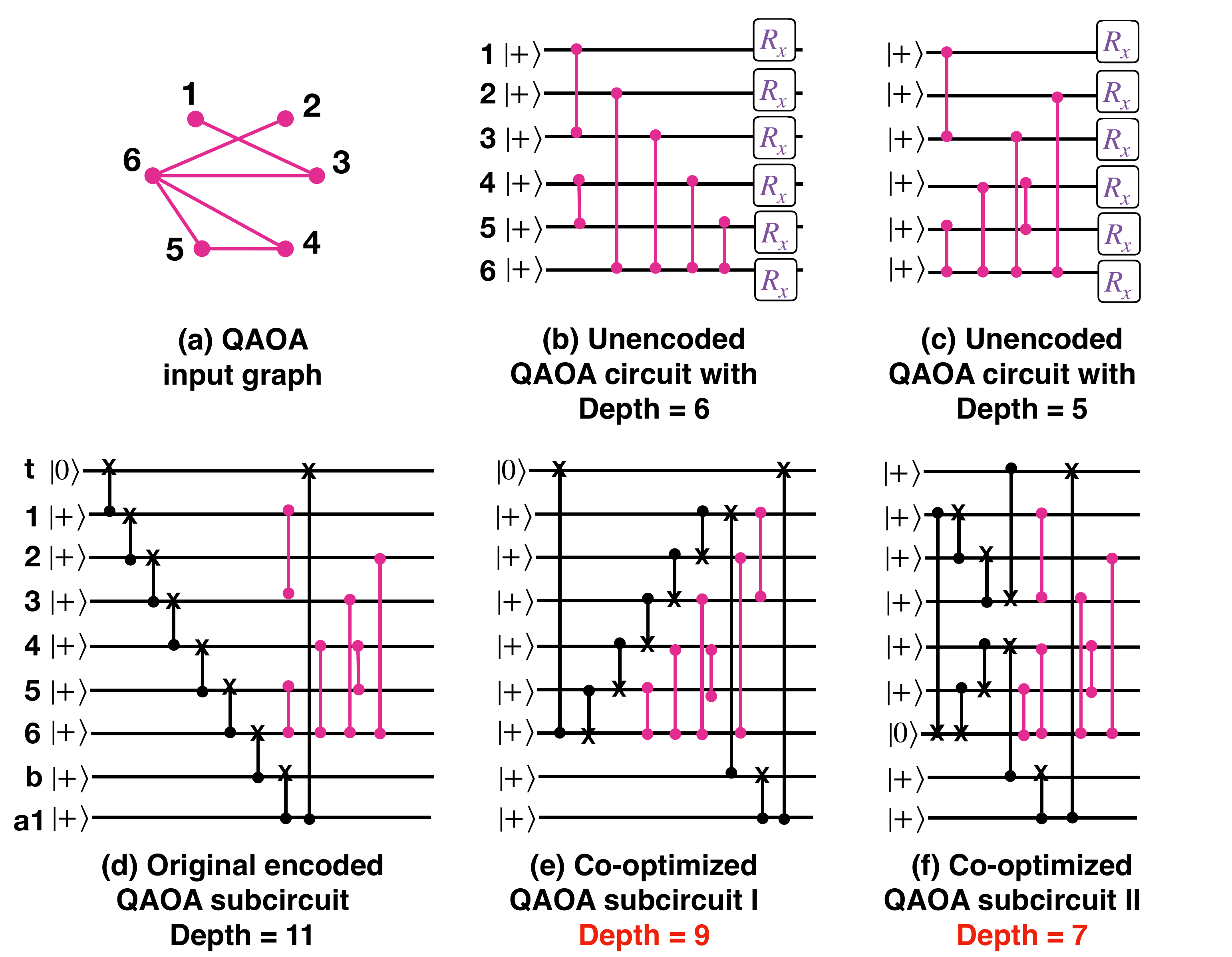}
\caption{
(a) QAOA input problem graph.
(b)-(c) Two valid QAOA circuits with different circuit depths for the same input problem in (a).
(d) Iceberg-encoded circuit using the original initialization gadget and the QAOA problem graph from (c).
(e) Optimized circuit with resynthesized initialization gadget.
(f) Further optimized initialization gadget with two branches, which reduces the circuit depth even more.
}
    \label{fig:gadgetOpt_qaoa}
\end{figure}

\begin{figure}[t]
    \centering
    \includegraphics[width=0.9\linewidth]{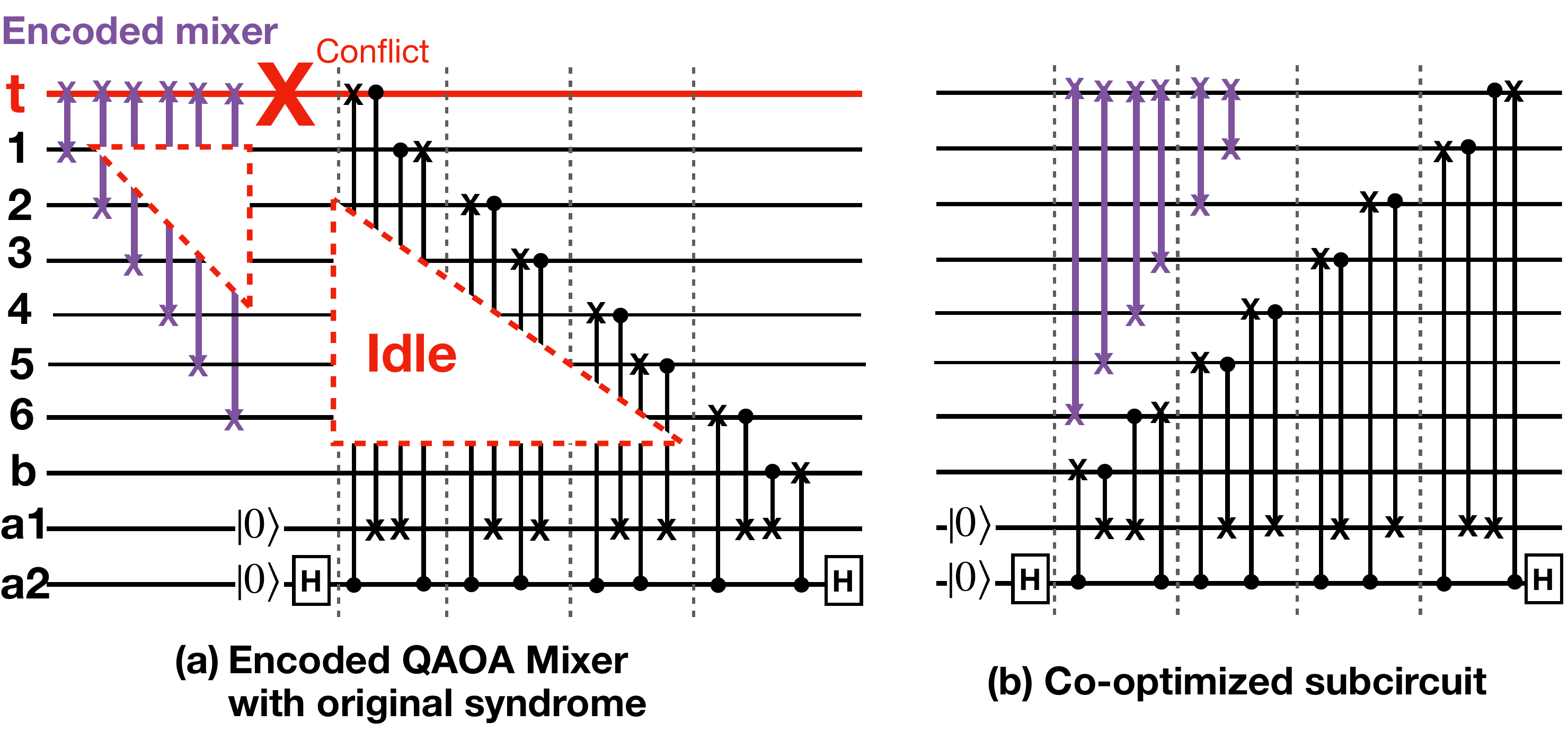}
    \caption{(a) An encoded QAOA mixer layer followed by a syndrome gadget. There is a huge idling area between these two components. 
    (b) Idling area reduced by co-compilation with the syndrome gadget resynthesis and gate commutation.}
    \label{fig:gadgetOpt}
\end{figure}

We begin by leveraging the flexibility in the qubit ordering of fault-tolerant gadgets. Typically, these gadgets are defined with an implicit qubit order $[t, 1, 2, \ldots, k, b]$ (see \Cref{fig:motivate_encoded_circuit}(c)), and their gates are arranged in a ``staircase'' pattern. Importantly, this qubit order can be arbitrarily permuted without impacting the gadget's functionality or fault tolerance.
For the initialization gadget, such reordering is possible because the output is a physical GHZ state~\cite{greenberger2007goingbellstheorem}, which is inherently symmetric under qubit permutations. For example, \Cref{fig:gadgetOpt_qaoa}(e) illustrates an initialization gadget with the qubit order $[t,6,5,4,3,2,1,b]$. By prioritizing frequently used qubits---such as qubit 6 in this case---the circuit depth can be reduced. In the given example, the optimized qubit order lowers the circuit depth from 11 to 9.
Furthermore, since the initialization gadget prepares a GHZ state, we can employ a two-staircase structure to halve the circuit depth (see \Cref{fig:gadgetOpt_qaoa}(f)). However, to preserve fault tolerance, the initialization cannot be split into more than two branches.
Additional splitting would necessitate extra ancilla qubits and measurements, thereby increasing resource overhead and circuit complexity.

The implicit order of the syndrome measurement gadget can also be modified since it measures the GHZ global stabilizers $S_z$ and $S_x$, which are invariant under permutations. \Cref{fig:gadgetOpt}(b) illustrates a syndrome measurement gadget with an implicit order of $[b, 6,5,4,3,2,1,t]$. Without resynthesizing the syndrome gadget, the qubit conflicts in the top qubit increase idling errors in the encoded mixer layer and the original syndrome gadget. Gadget resynthesis gives us an opportunity to allow the mixer layer to fully overlap with the new gadget (\Cref{fig:gadgetOpt}(b)). The implicit order of the final measurement gadget can also be changed as long as the classical post-processing is adapted accordingly. This flexibility is used in this work to reduce circuit overhead, specifically the circuit depth, leading to larger Iceberg code implementations with higher post-selection rates.

\subsection{Co-Compilation with Mixer Flexibility}
\label{subsec:co-compilation_with_mixer_flexibility}

The QAOA mixer consists of a layer of single-qubit $R_x$ rotation gates. In the unencoded circuit, these gates act independently on each qubit, so their ordering does not affect the circuit depth. However, when the circuit is encoded with the Iceberg code, each $R_x$ gate becomes a two-qubit $R_{xx}$ gate, introducing dependencies. Importantly, the order in which these $R_{xx}$ gates are applied remains flexible. By strategically reordering the gates in the encoded mixer layer, we can reduce the overall circuit depth.

Simply resynthesizing the gadget may not perfectly reduce circuit depth. As shown in \Cref{fig:gadgetOpt}, if only the syndrome gadget is flipped, a qubit conflict arises at qubit 6, and only two mixer gates can overlap with the syndrome gadget. However, by reordering the $R_{xx}$ gates, the mixer layer can fully overlap with the syndrome gadget, leading to greater depth reduction. Similar scenarios occur in \Cref{fig:gadgetOpt_qaoa}(e)-(f), where both gadget resynthesis and $R_{zz}$ gate reordering contribute to circuit depth reduction.

Gadget resynthesis and gate commuting alone are insufficient to achieve optimal circuit depth reduction. This motivates the design of a co-compilation framework that leverages both opportunities simultaneously. However, such flexibility causes the design space to grow exponentially with the number of qubits, necessitating a compiler that can efficiently capture and exploit these optimization opportunities.

\subsection{Leveraging Problem Symmetry}

\begin{figure}[t]
    \centering
    \includegraphics[width=0.45\linewidth]{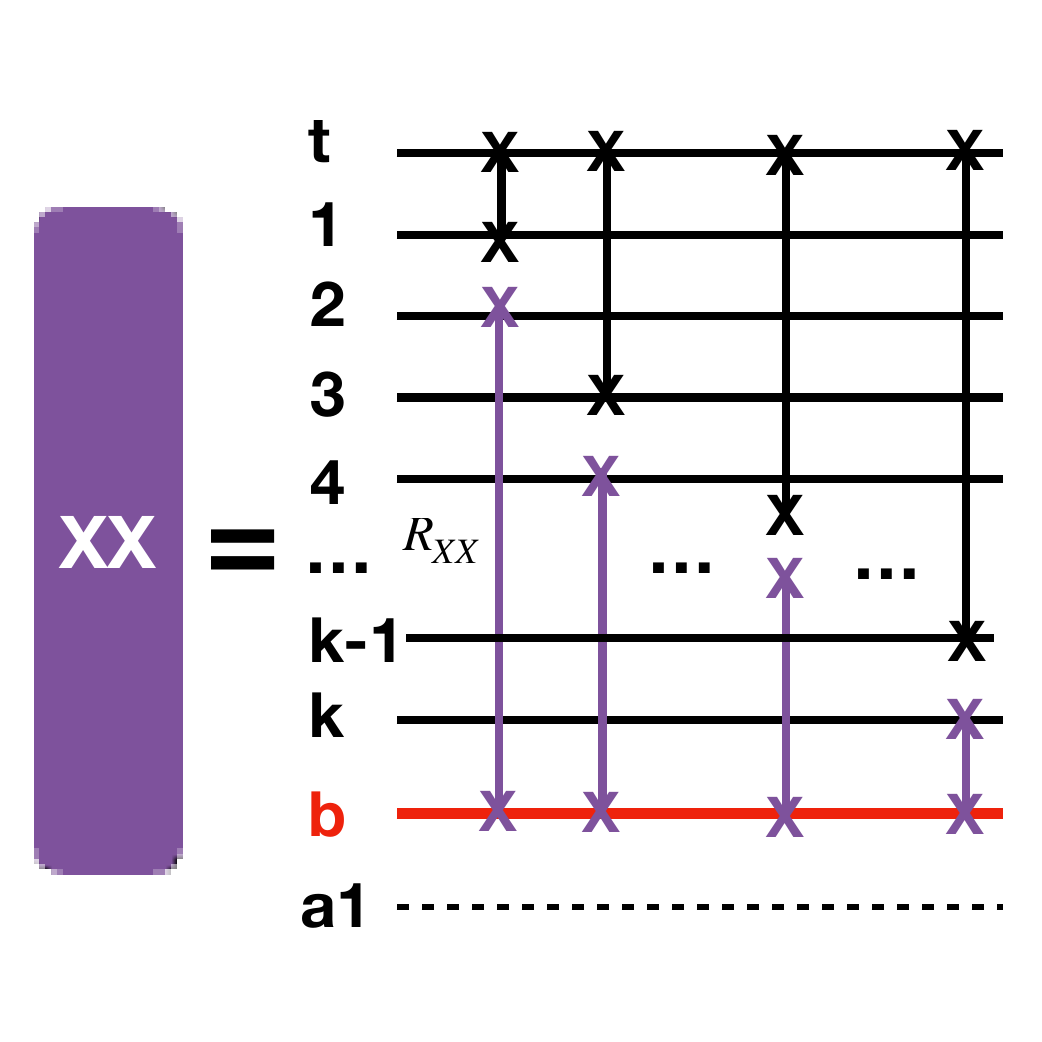}
\caption{ 
Optimized mixer implementation using the $\mathbb{Z}_2$ symmetry.  
The circuit depth of a mixer layer is reduced by half by implementing the $R_{xx}$ rotation, connecting half of the qubits to the bottom.  
For a detailed description, see \cref{subsec:co-compilation_with_mixer_flexibility}.
}
    \label{fig:symmetryOpt}
\end{figure}

We leverage the $\mathbb{Z}_2$ symmetry of the MaxCut QAOA circuits to further reduce circuit depth. In these circuits, flipping all qubits by the operator $X_1X_2\cdots X_k$ preserves the state. When encoded in the Iceberg code, we additionally have the symmetry imposed by the stabilizer operator $S_x$. Consequently, the Pauli operator $X_tX_b$ is also a symmetry of the encoded state, resulting in
\begin{equation}
    X_tX_i = X_t X_i \, S_x \, X_1X_2\cdots X_k = X_tX_i \, X_tX_b = X_b X_i .
\end{equation}
All logical rotations $\exp(-i\theta\overline{X}_i)$ on logical qubit $i\in[1, k]$ are implemented as physical rotations $\exp(-i\theta X_tX_i)$. This makes the implementation of the mixing operator of QAOA extremely non-parallel, as all physical rotations share the top qubit $t$. Thanks to the symmetries, the logical rotations can also be physically implemented as $\exp(-i\theta X_bX_i)$, allowing the use of the bottom qubit $b$ as a top qubit and thus halving the circuit depth of logical $\overline{X}$ rotations. A new circuit of a mixer Hamiltonian of QAOA is shown in \Cref{fig:symmetryOpt}. The proposed method exhibits a twofold reduction in circuit depth relative to prior work \cite{he2024performance}.

\section{Compilation Methodology }
Building on the optimization opportunities identified above, we now turn to the challenge of systematically exploiting the flexibility in both Iceberg gadgets and algorithmic circuit construction. The interplay between gadget synthesis and gate ordering dramatically expands the compilation design space, particularly as the number of logical qubits grows. While a straightforward approach might treat algorithm and gadget compilation as separate steps, such separation risks missing deeper co-optimization benefits. 

To address this, we use a graph representation to capture the qubit dependency but maintain the flexibility of circuit construction. Specifically, we use two graphs, $G_{uncompiled}$ and $G_{executable}$. We construct an uncompiled graph to estimate the depth of the uncompiled circuit by considering the flexibility of gadget construction and QAOA gate ordering together. The executable graph contains all the possible gates from both Iceberg gadgets and QAOA at each step, so we can explore all the valid executions easily at each step. Then, we introduce a tree search-based compilation framework over graphs that jointly optimizes the quantum error detection code and the algorithmic circuit. 
\update{The overall algorithm is summarized in Algorithm~\ref{alg:astar}.}
In the following, we detail how the search space is constructed and how the framework incorporates the flexibility and symmetries discussed previously.

\begin{algorithm}
\caption{\update{QAOA + Iceberg Circuit Co-compilation}}
\label{alg:astar}
\begin{algorithmic}[1]
\REQUIRE $ZZ\_graph$, $num\_syn$
\ENSURE Optimized Encoded QAOA Circuit

\STATE \textbf{Initialize:}
\STATE $top\_node \leftarrow {CreateNode}(n\_qubits, cost=0)$
\STATE $priority\_queue \leftarrow {PriorityQueue}()$
\STATE $priority\_queue.{push}(top\_node)$
\STATE \textbf{Searching:}
\WHILE{$top\_node.compiled\_gate\_count$ != $all\_gates$}
    \STATE $top\_node \leftarrow priority\_queue.{pop}()$

\STATE $(G_{exe}, G_{uncomp}) \leftarrow {GetRemainingGraphs}(top\_node)$

\STATE $candidates \leftarrow {FindMultipleMaxMatchings}(G_{exe})$

\FOR{ $gates$ in $candidates$}
    \STATE $child \leftarrow {CreateChild}(gates)$
    \STATE $child.parent = top\_node$
    \STATE $heuristic\_cost \leftarrow {EstimateCost}(G_{uncomp})$
    
    \STATE $child.cost \leftarrow top\_node.cost + 1$
    \STATE $child.score \leftarrow child.cost + heuristic\_cost$
    \STATE $priority\_queue.{push}(child)$
\ENDFOR

\ENDWHILE

\RETURN $top\_node$

\end{algorithmic}
\end{algorithm}

\subsection{Tree Search Framework}
The primary objective of our compilation is to minimize the idling area in an encoded circuit while preserving the fault-tolerance of the code gadgets. Equivalently, we aim to minimize the circuit depth.
We divide a quantum circuit into different layers such that gates within the same layer do not share qubits. Each layer then represents a distinct \textit{circuit state}, and the overall depth of the circuit corresponds to the total number of layers. 

We use \textit{nodes} to represent possible states of the quantum circuit in the search tree. Each node contains one possible layer of gates from both the algorithmic circuit and Iceberg gadgets. \textit{The source node} represents the initial state where nothing is compiled, while \textit{the goal node} represents the state where all gates are compiled. The compiler's task is to find the shortest path from the source node to the goal node.

The \textit{Expander} in the search framework broadens the search space from a given circuit state by generating the next layer of child nodes, each representing a possible valid circuit state in the subsequent step. These valid circuit states originate from the gate commutation in QAOA circuit and the flexibility inherent in the Iceberg code gadgets. 

All generated tree nodes are stored in a priority queue with priorities determined by a cost function. The search process involves repeatedly selecting the node with the minimum cost from the priority queue and generating the next layer of child nodes until the goal node is found in the queue. 

\subsection{Guidance in the Search}
\label{sec:costfunction}
The total number of gates in the circuit is finite; however, the number of valid circuit states 
grows exponentially in the number of layers. To efficiently traverse this vast search space, we employ a cost function $F(node)$, which estimates the optimal circuit depth by incorporating various optimization techniques. At a tree node, $F(node)$ is calculated as F(node) = G(node) + H(node), where $G(node)$ is the current cost, which is the path length from the source node to the current node. $H(node)$ is the heuristic cost that estimates the depth of the remaining uncompiled circuit. A more precise estimation leads to a more efficient search. With an admissible estimation, a lower bound of the length of the path from the current node to the goal node, the tree search framework guarantees an optimal search result \cite{alwin:qubitMapping, qubitMappingAstar, qaoaMappingAstar}.

In the Iceberg-encoded QAOA circuit, physical $R_{zz}$ gates, i.e., $\exp(-i\theta ZZ)$, commute with each other, as do physical $R_{xx}$, i.e., $\exp(-i\theta XX)$, gates. We use graphs to represent the QAOA phaser and mixer encoded gates, as well as the Iceberg code gadgets, to capture the inherent flexibility.

We design the heuristic depth estimation as:
\begin{equation}
depth\_est = \mathrm{max}_{v \in V}\left(\textstyle\sum_{(v,u) \in E} weight(v,u)\right),
\end{equation}
where $V$ and $E$ are the sets of vertices and edges in the uncompiled graph, respectively. The uncompiled graph $G_{uncompile}(V,E)$ is an aggregated representation of all operations from the uncompiled circuit at the current tree node. Each vertex in the uncompiled circuit represents a physical qubit, and each edge represents the interaction between two physical qubits. The weight of an edge indicates the number of two-qubit gates between the qubits. An example of an uncompiled graph is shown in \Cref{fig:uncompiledGraph}. 

It is straightforward to use a graph representation for the entangler layer and encoded mixer layer; however, the syndrome gadget is more complex. \Cref{fig:gadgetOpt} shows that all of the qubits in the syndrome gadget only have two consecutive operations on them, and both of them connect to ancilla qubits. So we construct an uncompiled graph for the syndrome gadget with all vertices connecting a vertex $a$, which represents two ancilla qubits, and the edge weight is 2. By adding subgraphs from each circuit component together, we can get an uncompiled graph and the depth estimation. In this example, vertex $a$ has the highest weight, and the estimated depth is 14.

\update{However, when nodes with different progress have the same depth estimation, the search loses efficiency in distinguishing promising paths. To address this issue, we introduce progress tracking into the cost function. 
We define the cost function by dividng depth estimation by the total number of compiled gates of the given node as follows:
\begin{equation}\label{eq:key}
H(node) = \frac{depth\_est}{total\_compiled\_gates}.
\end{equation}
This modification ensures that when two nodes have the same depth estimation, the one with more compilation progress (higher gate count) receives a lower effective cost and is prioritized. Similarly, when two nodes have the same progress, the node with the lower circuit depth estimation is expanded first. We present detailed scalability analysis in 
\Cref{sec:depth_analysis}.}

\subsubsection{Pre-processing}
\begin{figure}[t]
    \centering
    \includegraphics[width=0.9\linewidth]{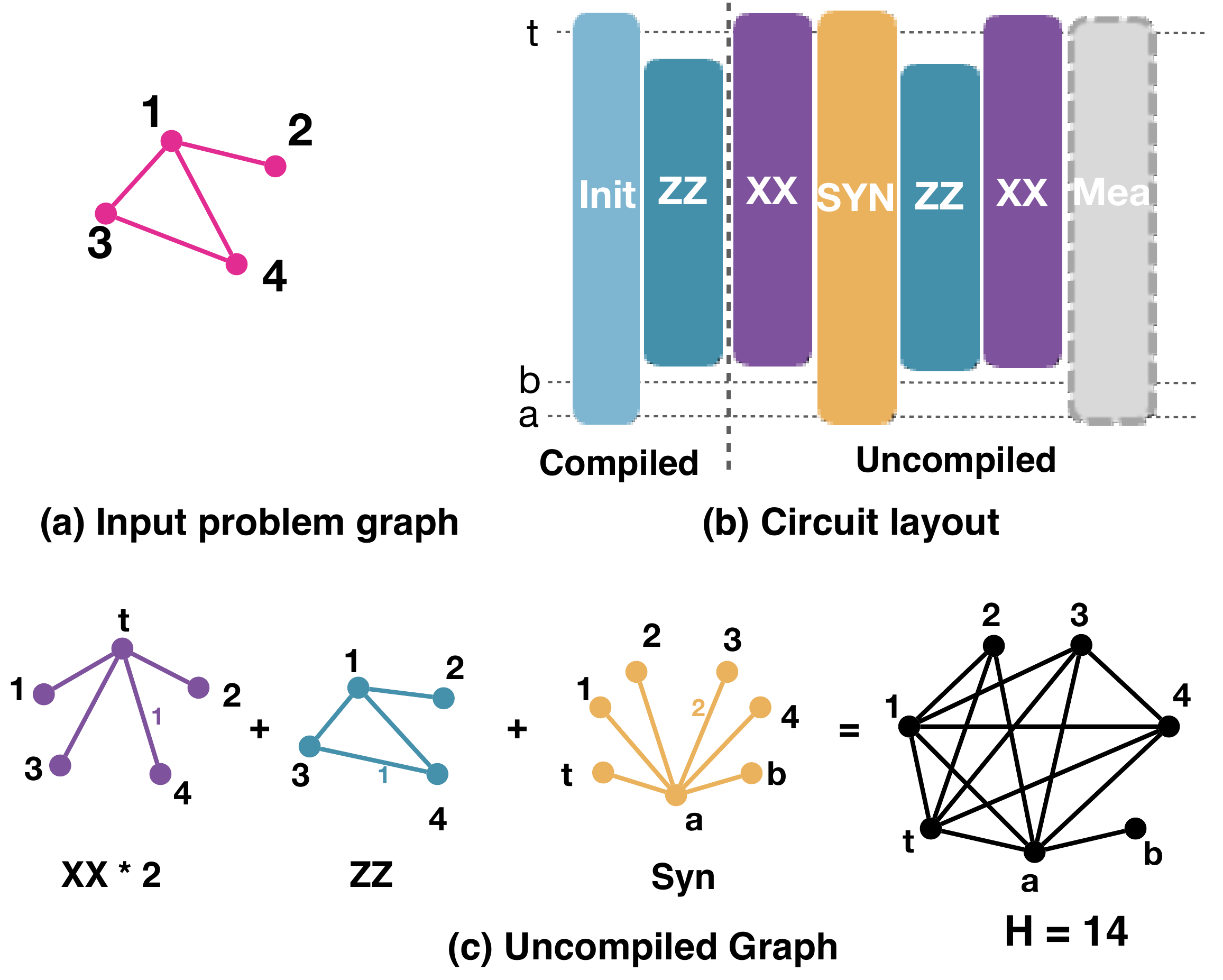}
    \caption{Uncompiled graph construction for the depth estimation. (a) The input QAOA problem graph with four vertices. (b) The predetermined Iceberg encoded circuit layout with one syndrome gadget. 
    (c) The uncompiled graph of the remaining uncompiled circuit. 
    Each subgraph stands for the operations in the corresponding circuit component. The number on each edge stands for the number of operations between two targets. The vertex $a$ has the largest weight, suggesting the depth estimation for the uncompiled graph is 14. 
    }
    \label{fig:uncompiledGraph}
\end{figure}

\ \\
 \quad\textbf{Predetermined Circuit Layout.}  To reduce the search space without compromising compilation quality, we predetermine the circuit layout. We define a circuit \textit{component} as a sequence of consecutive gates from the same gadget or the same QAOA layer. As illustrated in \Cref{fig:uncompiledGraph}(b), the encoded circuit begins with an initialization gadget and ends with a final measurement gadget, with one syndrome gadget placed in the middle.

\textbf{Predetermined Initialization Gadget.} An effective initialization can fully overlap with the first ZZ component. However, the advantages of optimizing the initialization gadget are limited, particularly when the circuit is deep. Therefore, we predetermine the structure of the initialization gadget based on the degree of the vertices in the QAOA input problem graph. Qubits corresponding to high-degree vertices are prioritized in the gadget construction. For instance, in \Cref{fig:gadgetOpt_qaoa}(a), vertex 6 has a higher degree than the other vertices, so it is connected by CNOT gates earlier than the others in the initialization gadget. The compiled circuit has shorter depth in \Cref{fig:gadgetOpt_qaoa}(e). 

\textbf{Final Measurement.} In the circuit depth estimation, the circuit with fewer components is easier to estimate. So we temporarily remove the final measurement gadget in the depth estimation. We append it to the circuit once all gates before the final measurement are compiled heuristically.

 \begin{figure}[t]
    \centering
    \includegraphics[width=0.9\linewidth]{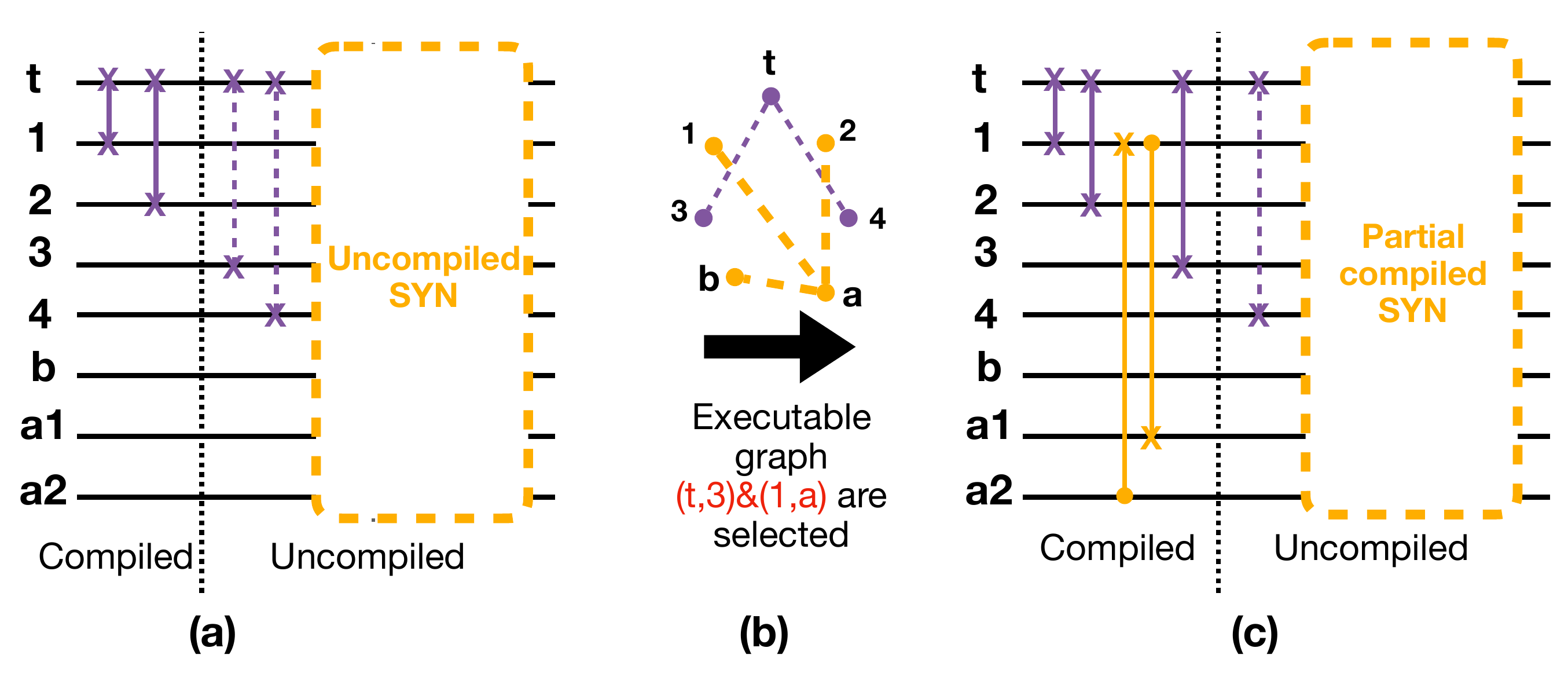}
\caption{
Search space expansion with a syndrome gadget for a given circuit status. (a) A partial compiled mixer layer followed by a not compiled syndrome gadget. All possible valid syndrome gates will be considered during expansion. (b) Graph representation of all possible executable gates in the next step. (c) Gate $R_{xx}(t,3)$ and two syndrome gates corresponding to the edge $(1,a)$ are chosen in the expansion.
}
    \label{fig:synExpand}
\end{figure}

\subsection{Expanding the Search Space}
In our tree search framework, the searching and search space construction happen simultaneously. At each step, we will find the node with the minimum cost and expand the next generation of child nodes. The question is how to choose gates properly for the child nodes, including in the case when the compilation involves a syndrome gadget. 

Before we expand the search space, we want to know all the possible valid gates for the next step. Then, we can choose different combinations of those gates that can be beneficial to the circuit depth reduction.  Gates in each circuit component do not have strict time dependency; however, circuit components do have. We define the gate dependency constraint across different circuit components. 

\textbf{Constraint 1. } If a gate $g(a,b)$ in the circuit component $i$ is executed, then all gates in the circuit component $j$ involving qubit $a$ and $b$ have to be finished, where $i > j$.  

For example, the $R_{xx}(t, i)$ gate in one QAOA layer has to wait until all $R_{zz}$ gates involving qubit $i$ finish first in the same QAOA layer. 

Due to the flexibility in QAOA circuit and Iceberg QED, we still use a graph to represent all possible gates that we can compile at the next step. With the constraint discussed above, we define an $G_{executable}(V,E)$, where $V$ contains all the physical qubits and $E$ stands for a set of edges that are related to executable gates from the uncompiled circuit. To construct such an $G_{executable}(V,E)$, we start from the first circuit component that is not fully compiled and a list of all physical qubits $l$. We add edges corresponding to those uncompiled gates to the graph in the current circuit component, and remove corresponding occupied qubits in $l$ to comply with constraint 1, because they will not be used by the following circuit. If there are unoccupied qubits remaining in $l$, we check the next circuit component and repeat until we use all qubits or no more edges can be added to the graph. The subgraphs from different circuit components are independent. Then we can simply choose different combinations of edges to proceed with the compilation.

For example, assuming all gates before the mixer component in \Cref{fig:synExpand}(a) are compiled and two gates in this component are also compiled. So qubits $t$, 3, and 4 will be occupied by this component, and edges $(t,3)$ and $(t,4)$ will be added to $G_{executable}$ (the purple subgraph in \Cref{fig:synExpand}(b)). %

Next, the list $l$ still contains qubit 1,2, and $b$. So we can use them to compile gates from the next circuit component, which is a syndrome gadget in this case. Because each qubit in the syndrome gadget connects to two ancilla qubits, and which qubit is used first does not matter. We could connect each available vertex to the vertex $a$ and add this subgraph to the $G_{executable}$. An example is shown in \Cref{fig:synExpand}(b), the subgraph in yellow stands for the valid choices at the next step. %
Since the next syndrome gadget will use all qubits, the list $l$ will be empty, and no more edges can be added to $G_{executable}$.

The last step is how to choose edges from the executable graph and proceed with the compilation. Ideally, we can try all possible gate combinations and each combination stands for a child node. However, if too many child nodes are generated, the search space grows exponentially, and the compilation time would be intractable. So we use the maximum weight matching algorithm~\cite{maxWeightMatching} to heuristically pick a few combinations in the executable graph. 
One example is in \Cref{fig:synExpand}(b)-(c). We choose $(t,3)$ and $(1,a)$ from the executable graph and add the corresponding gates back to the circuit. Note that the gates in the syndrome gadget have to follow the pattern of the Iceberg code to preserve fault tolerance. 

\begin{figure}[t]
    \centering
\includegraphics[width=0.9\linewidth]{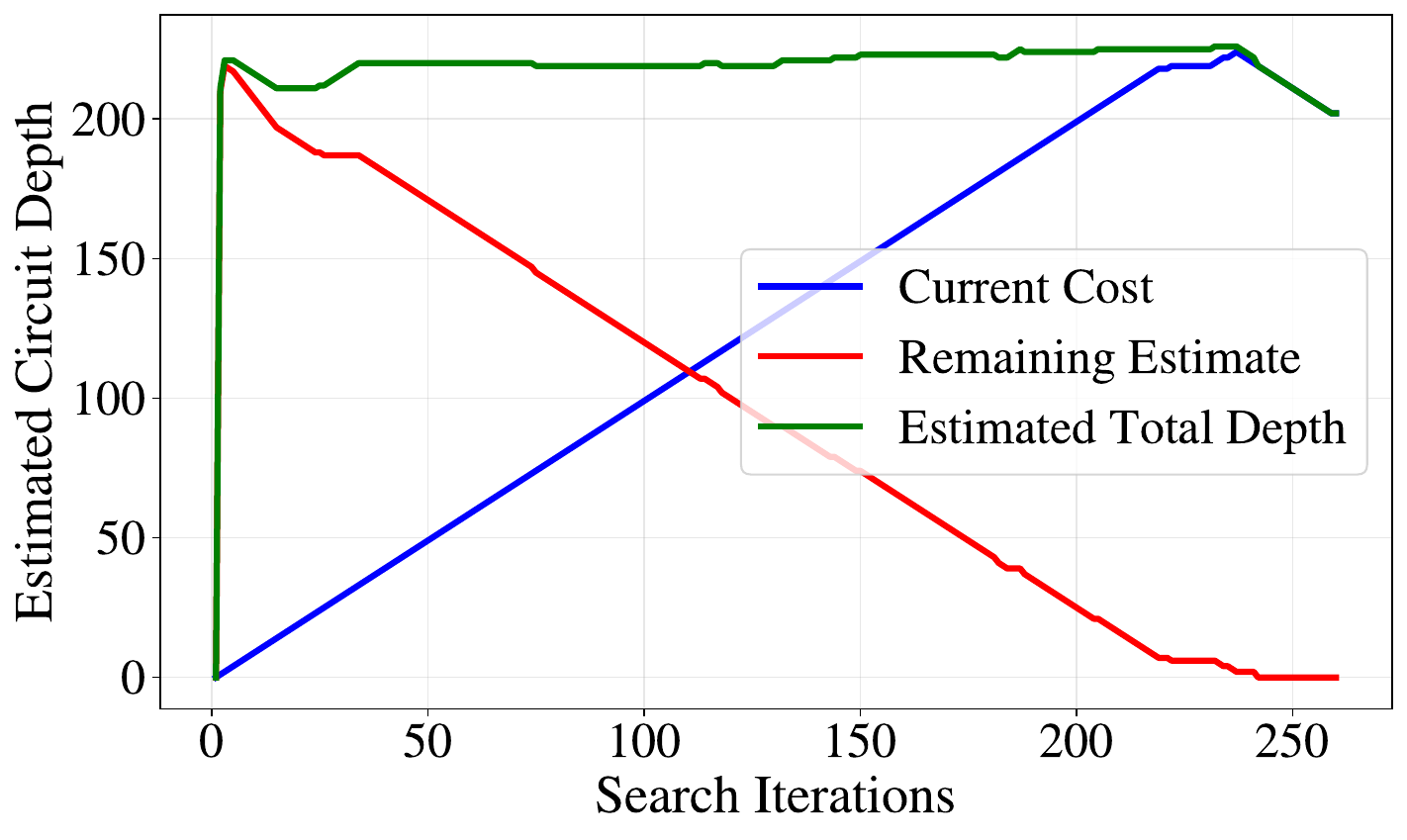}
    \caption{
    \update{The dynamic behavior of the search algorithm. We track the dynamics of quantum circuit compilation for a representative 22-qubit random graph instance with density 0.8 and expansion size 10. %
    }}
    \label{fig:search_progress}
\end{figure}

\update{
\subsection{Robustness Analysis}\label{sec:robustness}

To analyze the robustness of the proposed search method, we examine how the estimated circuit depth evolves during search iterations. We find that the circuit depth estimation remains stably close to the final compiled circuit depth, indicating that the degree-based estimation is surprisingly accurate.

\Cref{fig:search_progress} illustrates the dynamic behavior of the proposed search algorithm during quantum circuit compilation for a representative 22-qubit random graph instance with density 0.8 and expansion size 10 (10 child nodes in each expanding). We track the metrics of each node popped from the priority queue during search iterations. The x-axis represents the search iteration, while the y-axis shows the estimated circuit depth at each iteration. Three key metrics track the search progression: the current cost (blue line) represents the circuit depth achieved at the current search node, the remaining estimate (red line) shows the heuristic prediction of additional depth required to complete compilation, and the estimated total depth (green line) represents their sum, which guides node selection in the priority queue. The stable convergence pattern demonstrates the accuracy of our heuristic function throughout the search process.

}

\begin{figure*}[t]
    \centering
\includegraphics[width=1\linewidth]{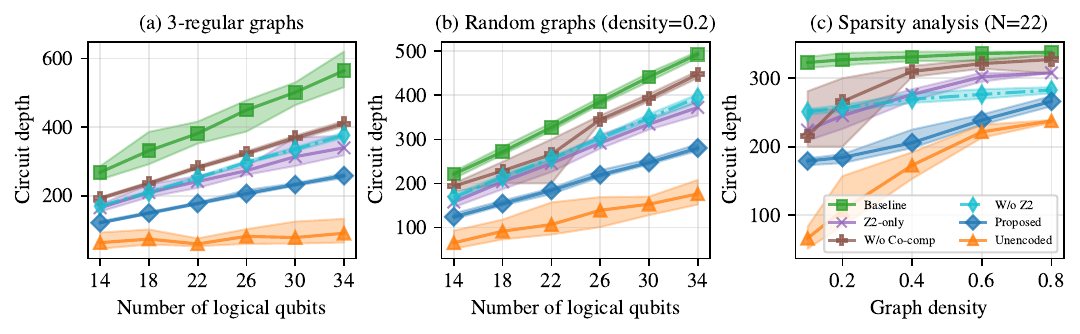}
    \caption{
    Depth of the unencoded circuit and the encoded circuit with different optimizations. 
    (a) Compiled two-qubit depth for 3-regular graphs with varying numbers of algorithmic qubits. The proposed pipeline achieves an average of $54\%$ improvement over the unoptimized Iceberg circuits.
    (b) Depth for random graphs with a density of 0.2. The proposed pipeline achieves an average of $43\%$ improvement over the unoptimized Iceberg circuits.
    (c) With varying densities of random graphs, the proposed pipeline consistently outperforms the others. The improvement diminishes as the graphs become denser, as there is less room for improvement. 
    The shaded regions represent the minimum and maximum circuit depths across 10 instances.
    }
    \label{fig:ablation_depth}
\end{figure*}
\section{Numerical Results of QAOA}

Experimental results are presented in this section, focusing on two types of graphs: 3-regular graphs and Erdős--Rényi random graphs. 
The compiler is evaluated using various optimization techniques, and comparisons are made between different compiler configurations. 
Additionally, the encoded circuit is analyzed to gain deeper insights into the QAOA algorithm and to identify issues related to the Iceberg code.

\subsection{Experiment Setup}

\textbf{Backend.} We conducted all experiments on the Quantinuum H2-1 machine~\cite{H2, Moses2023}  and its emulator. The H2-1 device has 56 physical qubits with all-to-all connectivity, 99.997\% single-qubit gate fidelity, and 99.87\% two-qubit gate fidelity.
Unless otherwise noted, experiments with $\leq$ 26 qubits are conducted in the emulators via the Quantinuum cloud service. \update{The H-series emulators perform state-vector simulation with noise inserted into the circuit to mimic characteristics of real devices. For more detailed emulator data, please refer to~\cite{quantinuum_h2_emulators}.}

\textbf{Metrics.} We report the circuit \emph{depth} of two-qubit gates to compare our method with other methods. Circuits with lower depths suffer less coherent error and have better fidelity. Since the H2-1 backend has all-to-all connectivity, there is no SWAP gate introduced in the compilation. All methods have the same gate count, so we do not report it. The \emph{post-selection rate} is an important metric to study the overhead of Iceberg code. It is the ratio of remaining shots to the total completed shots; a higher ratio is preferable. If the post-selection rate is too low, then the result would be less reliable. 

We refer to the negative cut value of a bitstring as energy. To evaluate the performance of the QAOA algorithm in the MaxCut problem, we use the success probability (the probability associated with bitstrings having the lowest energy) and the \emph{approximation ratio} (AR), $\alpha(\psi) = (\vert E \vert - \langle \psi|C|\psi\rangle)/{2f_{max}}$, where $\vert E \vert$ is the number of edges in problem graph and $f_{max}$ is a precalculated optimal cut value of this graph.

\textbf{Benchmarks.} All graph instances are generated using the Python library \texttt{NetworkX}~\cite{hagberg2020networkx} with random seeds. We prepared two types of graphs: regular graphs with a vertex degree of 3, and random graphs with densities of 0.1, 0.2, 0.4, 0.6, and 0.8. In the circuit depth analysis, each data point is based on ten graph instances to minimize statistical errors.

Throughout the experiments with encoded circuits, the syndrome gadgets are evenly distributed in the circuit such that algorithmic computational gates are partitioned into chunks with a similar number of gates.

\begin{figure*}[t]
    \centering
\includegraphics[width=1\linewidth]{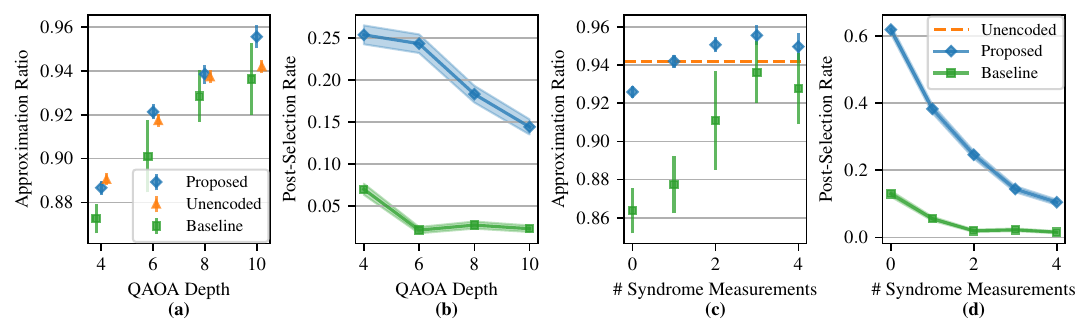}
    \caption{QAOA performance \update{on H2-1 emulator} for MaxCut on a $k=22$ 3-regular graph instance with varying QAOA depth $p$ and numbers of syndrome measurements. (a)-(b) The encoded circuits include three syndrome measurements. (c)-(d) The encoded circuits have QAOA depth p = 10.  The co-optimized circuit consistently outperforms the baseline circuit in both the approximation ratio and post-selection rate. The error bars and shaded regions represent the standard error.
    }
    \label{fig:QAOA_depth_syn}
\end{figure*}

\subsection{\update{QAOA Depth Evaluation}}
\label{sec:depth_analysis}

\begin{table}[t]
\centering
\begin{adjustbox}{width=\linewidth}
\begin{tabular}{c|ccccc}
\hline
Technique  & Baseline & $\mathbb{Z}_2$-only & W/o Co-comp & W/o $\mathbb{Z}_2$ & Proposed \\
\hline
Co-compilation & - & - & - & \checkmark & \checkmark\\
\hline
$\mathbb{Z}_2$ symmetry & - & \checkmark & - & - & \checkmark\\
\hline
Tree search & - & - & \checkmark & \checkmark & \checkmark\\
\hline
\end{tabular}
\end{adjustbox}
\caption{\update{Techniques used in the various compilation workflows in \Cref{fig:ablation_depth}.}}
\label{tab:ablation}
\end{table}

In this section, we present comparisons of different compilation passes with different benchmarks. The number of QAOA layers is set to 10, and if the circuit is encoded, it includes 3 syndrome gadgets. \update{\Cref{tab:ablation} summarizes the techniques used in these compilation passes, compared in \Cref{fig:ablation_depth}.} \textit{Proposed} refers to our compiler with all proposed techniques applied, \textit{W/o $\mathbb{Z}_2$} shows the results without leveraging the symmetry property, \update{\textit{W/o Co-comp} refers to separating algorithm and Iceberg gadgets with barriers and performing tree-search optimization on individual partitions, and \textit{$\mathbb{Z}_2$-only} refers to utilizing the $\mathbb{Z}_2$ symmetry to optimize the mixer circuit depth in the baseline but does not apply other co-optimization techniques}.

In \Cref{fig:ablation_depth} (a) and (b), we show that our compiler achieves better circuit depth compared to the baseline version of the Iceberg-encoded circuit. With only the gadget resynthesis, the circuit depth is reduced by up to $36.6\%$ for regular graphs and $23.4\%$ for random graphs. 
By studying the problem's $\mathbb{Z}_2$ symmetry, we can further reduce the circuit depth through algorithm and error detection code co-optimization. \update{Compared against the baseline,} the depth is reduced by up to $54.8\%$ for regular graphs and $43.9\%$ for random graphs with density 0.2. 
\update{Compared against the $\mathbb{Z}_2$-only results, the depth is reduced by up to $30\%$ for regular graphs and $28.2\%$ for random graphs with density 0.2. Compared against the \textit{W/o $\mathbb{Z}_2$} results, the depth is reduced by up to $37\%$ for regular graphs and $37.4\%$ for random graphs with density 0.2.}

In \Cref{fig:ablation_depth} (c), we show circuit depth analysis with different graph sparsity. We fix the number of logical qubits to 22 and insert three syndrome gadgets in the encoded circuits. The logical circuit depth increases as the graph density increases. 
However, the circuit depth of the baseline circuits remains relatively stable. 
\update{
Because of the top-qubit conflict between mixers and syndrome gadgets, there is barely any gate parallelism among those components, and the idling area is large. 
The additional $R_{zz}$ gates for denser graphs can be filled into this idling area. 
}
However, with gadget resynthesis, we could approximately reduce the circuit depth of 100 constantly, which matches the circuit depth of $5.5k + O(1)$ from five gadgets. With symmetry optimization, we can further reduce the circuit depth and the idling area in the circuit. \update{For both the baseline and the proposed method, the improvement of $\mathbb{Z}_2$ vanishes as density increases, because the mixer gates can be more easily merged into the phaser layer.}

\subsection{QAOA Performance Analysis}

In \Cref{fig:QAOA_depth_syn}(a)-(b), we show the approximation ratio and the post-selection rate as a function of QAOA depth $p$ for a fixed 22-qubit instance executed on the H2-1 emulator. \update{Each emulator simulation takes 5000 shots.} We compare the performance of the original unencoded QAOA, the Iceberg-encoded QAOA with the baseline compilation, and with our compilation. For encoded circuits, we fix the number of syndrome measurements to be three. We observe that our compilation leads to significantly higher approximation ratios compared to the baseline compilation. We produce considerably shallower circuits, which incur less exposure to hardware noise. After post-selection, our AR is slightly higher than that of the unencoded circuit, except for $p=4$, where three syndrome measurements bring more overhead than benefits.

\Cref{fig:QAOA_depth_syn}(c)-(d) show the same comparisons with different numbers of syndrome measurements and a fixed QAOA depth of 10. As expected, the post-selection rate drops as the number of syndrome measurements increases, indicating more errors are detected. However, the syndrome measurement gadget has a nontrivial overhead on circuit depth, and more syndrome measurements also mean more exposure to errors 
that can accumulate to cause undetected logical errors. 
We see that for 10-layer QAOA, three syndrome measurements lead to the highest approximation ratio.%

\update{
\subsection{Scalability and Hyperparameter Study}\label{sec:scalability}

\begin{figure}[t]
    \centering
\includegraphics[width=1\linewidth]{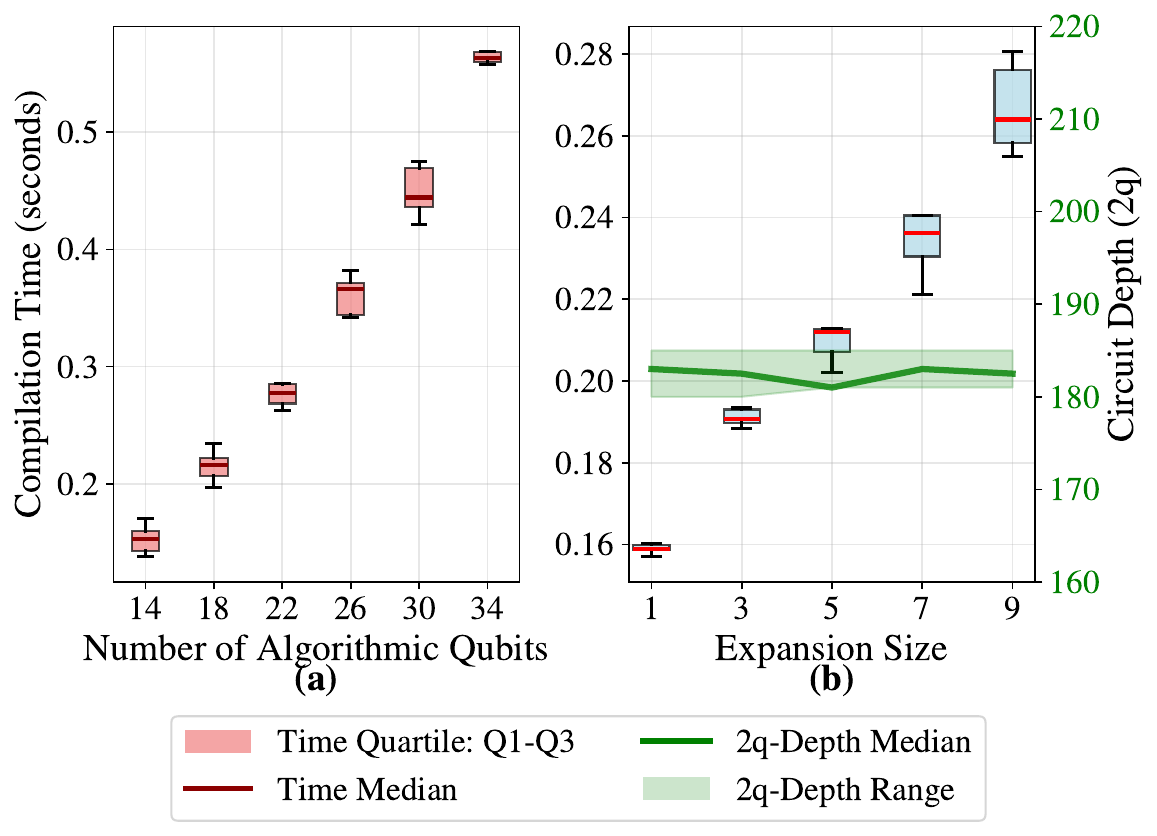}
\caption{
\update{Scalability and hyperparameter study of the search algorithm. 
(a) Compilation time as a function of number of algorithmic qubits where each data point represents the statistical distribution from 10 independent runs using regular degree-3 graphs with expansion size 10. (b) Trade-off between expansion size and both compilation time and final 2-qubit gate depth for $k=22$ instances. The box plots represent time quartiles where the box boundaries indicate the first and third
quartiles (Q1--Q3), the red line shows the median compilation time, and whiskers extend to the minimum and maximum observed values across runs. The green line with shaded region displays the median 2-qubit circuit depth and its variability range (min--max) across different expansion sizes.}}\label{fig:scalability_analysis}
\end{figure}

We conduct scalability and hyperparameter studies on 10-layer QAOA circuits with 3 syndrome gadgets and random 3-regular graphs. \Cref{fig:scalability_analysis} (a) shows that the compilation time of our proposed search algorithm scales linearly in the problem size, enabling scalable compilation for production-scale problems we hope to solve in the future. The actual time taken is within a second for all instances we ran, demonstrating practically efficient compilation. We also show the quartiles across 10 random instances for each problem size and observe that the fluctuations are within 15\%. The scaling of the fluctuations is also linear in the problem size, further indicating the robustness of our method.

\Cref{fig:scalability_analysis} (b) shows the compilation times and compiled depths of ten 22-qubit instances using varying expansion sizes in the search algorithm. We see that increasing the expansion size only incurs a linear cost to the compilation time, making it affordable to expand the exploration space. Nonetheless, as we have discussed in \Cref{sec:robustness}, our depth estimation heuristic is remarkably accurate, enabling the algorithm to produce high-quality results even with just single-node expansions. For the subsequent experiments in this paper, we set the expansion size to be one.}

\section{Break-even Performance on Hardware}

In the previous section, we showed promising results in circuit depth reduction and QAOA performance across different scenarios on the H2-1 emulator. Here, we present the performance results on the H2-1 hardware. 
\update{All circuits executed on Quantinuum hardware or its emulator are compiled again by \texttt{tket}~\cite{sivarajah2020t} to convert gates to the basis gates of Quantinuum devices. In this process, we set the optimization level to 2 to enable the gate commutation optimization for the QAOA circuit. Note that level-2 optimization will apply gate cancellation to the circuit such that some gates will be cancelled and the fault-tolerant property of Iceberg will be broken. For example, two CNOT gates that interact with the two ancilla qubits in the final measurement will be cancelled. To avoid that, we insert barriers to each gate in the gadget, then compile it with the \texttt{tket} compiler. All barriers are removed before the circuit execution. Each hardware circuit takes 2000 shots due to the limited hardware resources}

\begin{figure*}[t]
    \centering
\includegraphics[width=1\linewidth]{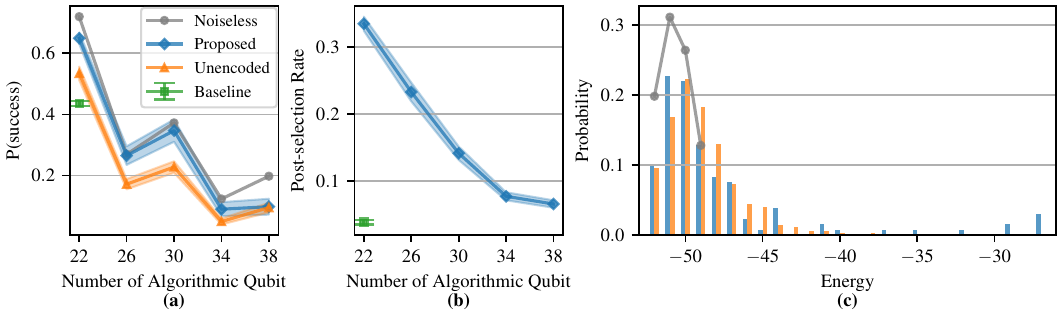}
    \caption{New state-of-the-art hardware results on Quantinuum H2-1. (a) The $p=10$ QAOA success probability is shown with varying numbers of algorithmic qubits. The proposed pipeline achieves beyond-break-even performance with up to $k=34$ algorithmic qubits, while performance is indistinguishable at $k=38$ qubits. The previous Iceberg QAOA (green dot) shows a lower success probability than the unencoded circuit at $k=22$~\cite{he2024performance}. All Iceberg QAOA circuits include three syndrome measurements.
    (b) %
    The proposed pipeline increases post-selection rates from $4\%$ to $33\%$ at $k=22$, and still achieves a $6.6\%$ post-selection rate at $k=38$. Shaded areas in (a) and (b) show standard error. 
    (c) The energy probability associated with the $k=38, p=10$ QAOA circuit. The Iceberg and unencoded circuits exhibit similar performance, as shown in (a), while the energy distributions associated with the Iceberg circuit display a long-tail behavior.
    }
    \label{fig:hardware_results_panel}
\end{figure*}

In \Cref{fig:hardware_results_panel}(a), we compare the success probability of QAOA on 3-regular graphs between noiseless, proposed co-compiled, unencoded, and baseline circuits. All QAOA circuits have $p=10$, and the encoded circuits include three syndrome measurements. In the previous experiment using state-of-the-art compilation~\cite{he2024performance}, Iceberg QAOA performed worse than the unencoded circuit at $k=22$. However, with the improved compilation circuit, Iceberg QAOA outperforms the unencoded version at $k=34$ and remains competitive at $k=38$. Additionally, its success probabilities are very close to the noiseless results up to $k=30$. The noiseless success probability for large $k$ QAOA is calculated using exact tensor network contraction~\cite{gray2021hyper}. 
Enabled by the proposed co-compilation pipeline, we have extended the break-even point of QAOA on 3-regular graphs from $k=20$ to $k=34$. The post-selection rates are shown in \Cref{fig:hardware_results_panel}(b). As expected, the post-selection rates of our co-compiled circuits decrease as the circuit size increases. However, the post-selection rate at $k=38$ remains manageable, around $6.6\%$. Notably, we improved the post-selection rate from $4\%$ to $33\%$ at $k=22$ as compared to \cite{he2024performance}. 

The hardware progress evident from our results is of independent interest. Specifically, we use the same algorithmic circuits for $k=26$ and $k=30$ as in the H2-1 experiments reported in Ref.~\cite{Shaydulin2023npgeq}. While Table 2 of Ref.~\cite{Shaydulin2023npgeq} reports unencoded success probability of $\approx 0.1$ for both $k=26$ and $k=30$, we observe the unencoded value of $0.17\pm 0.02$ and $0.23\pm 0.02$ respectively (errors are standard error). This improvement shows the reduction in H2-1 error rates between the summer of 2023 and the spring of 2025. Furthermore, protecting the circuit with Iceberg code improved these values to $0.26\pm 0.02$ and $0.35\pm 0.04$, respectively, to within the standard error of the noiseless values of $0.27$ and $0.37$.

Additionally, we present the probability distribution of the obtained energies associated with post-selected QAOA samples at $k=38$ (\Cref{fig:hardware_results_panel}(c)). We report only the probabilities of the lowest four energies of the noiseless simulation result, as calculating the full distribution at $k=38$ is computationally expensive, and our focus is on the behavior of high-quality solutions. While the Iceberg and unencoded circuits exhibit similar success probabilities, the probability of sampling the lowest two energies is higher in the encoded circuit than in the unencoded one. Notably, the Iceberg QAOA displays a longer-tail energy distribution compared to the unencoded circuit. 
This is because, as the number of logical qubits grows, undetected errors are also more likely to produce high-weight logical errors. 
For example, a weight-2 error $X_tX_i$ affects a single logical qubit $i$, while another weight-2 error $Z_tZ_b$ (corresponding to the global logical operator $\overline{Z_1Z_2\cdots Z_k}$) affects all logical qubits. Combining this high chance of global errors with the local nature of the MaxCut Hamiltonian, we speculate that likely local errors in the Iceberg code produce large energy shifts that form such a long tail in the energy distribution.

\section{Compilation of Circuits Beyond QAOA}

Beyond QAOA, the innovations in error detection, circuit compilation, and resource optimization presented in this work are broadly applicable to a wide range of quantum algorithms. 
In particular, we also consider the impact of these techniques on other important algorithmic primitives, such as the quantum Fourier transform (QFT) \cite{shor1994algorithms,nielsen2000quantum} and instantaneous quantum polynomial (IQP) \cite{bremner2011classical,shepherd2009temporally} circuit. 
These benchmarks are widely used to characterize the performance and scalability of quantum hardware and software stacks. 
By extending our analysis to these additional workloads, we demonstrate that the benefits of efficient error detection and circuit optimization are not limited to a single algorithmic family, but can enhance the practical feasibility and fidelity of diverse applications.

\Cref{tab:iqpQFT} presents the circuit depth of Iceberg-encoded circuits with and without compilation. 
To encode the QFT and IQP circuits, the original circuits are first decomposed into rotation gates using Qiskit~\cite{qiskit}.
The decomposed circuits are then encoded, with one or two syndrome detection steps inserted evenly throughout. 
The IQP circuit consists of $\log(k)$ layers, each comprising a layer of single-qubit rotation gates and a layer of $R_{zz}$ gates, where $k$ is the number of logical qubits. Optimization of the IQP circuit can reduce the circuit depth by up to 37\%. The gadget rate is defined as the ratio of the depth contributed by gadgets to the baseline circuit depth, serving as an indicator of the potential for co-optimization.
QFT circuits exhibit a relatively low gadget rate, which inherently limits the scope for optimization. This is primarily due to the all-to-all connectivity and the structure of QFT circuits, which are composed mainly of $H$ and controlled rotation gates that are relatively costly in the Iceberg encoding. However, for small instances, our approach still achieves a depth reduction of up to 13\%.

\begin{table}[ht]
\centering
\begin{adjustbox}{width=1\linewidth}
\begin{tabular}{|c|c|>{\columncolor{gray!10}}c|>{\columncolor{gray!10}}c|>{\columncolor{gray!10}}c|>{\columncolor{gray!20}}c|>{\columncolor{gray!20}}c|>{\columncolor{gray!20}}c|}
\hline
\multirow{3}{*}{\textbf{Circuit}} & \multirow{3}{*}{\textbf{Metric}} 
    & \multicolumn{3}{c|}{\cellcolor{gray!30}\textbf{1 Syndrome}} 
    & \multicolumn{3}{c|}{\cellcolor{gray!50}\textbf{2 Syndromes}} \\
\cline{3-8}
 & & \multicolumn{6}{c|}{\textbf{Algorithmic Qubits}} \\
\cline{3-8}
 & & \textbf{12} & \textbf{20} & \textbf{30} & \textbf{12} & \textbf{20} & \textbf{30} \\
\hline
\multirow{4}{*}{IQP} & Baseline        & 83  & 143 & 217 & 90  & 168 & 242 \\
\cline{2-8}
                     & Optimized       & 52  & 102 & 149 & 65  & 128 & 173 \\
\cline{2-8}
                     & Depth Reduction & 37\% & 29\% & 31\% & 28\% & 24\% & 29\% \\
\cline{2-8}
                     & Gadget Rate     & 0.55 & 0.49 & 0.46 & 0.72 & 0.58 & 0.57 \\
\hline
\multirow{4}{*}{QFT} & Baseline        & 214 & 485 & 830 & 231 & 507 & 846 \\
\cline{2-8}
                     & Optimized       & 190 & 446 & 777 & 202 & 466 & 797 \\
\cline{2-8}
                     & Depth Reduction & 11\% & 8\% & 6\% & 13\% & 8\% & 6\% \\
\cline{2-8}
                     & Gadget Rate     & 0.22 & 0.14 & 0.12 & 0.28 & 0.19 & 0.16 \\
\hline
\end{tabular}
\end{adjustbox}
\captionsetup{font=small}
\caption{Two-qubit depth of Iceberg-encoded QFT and IQP circuits for different number of algorithmic qubits.}
\label{tab:iqpQFT}
\end{table}

\begin{figure}[t]
    \centering
\includegraphics[width=0.9\linewidth]{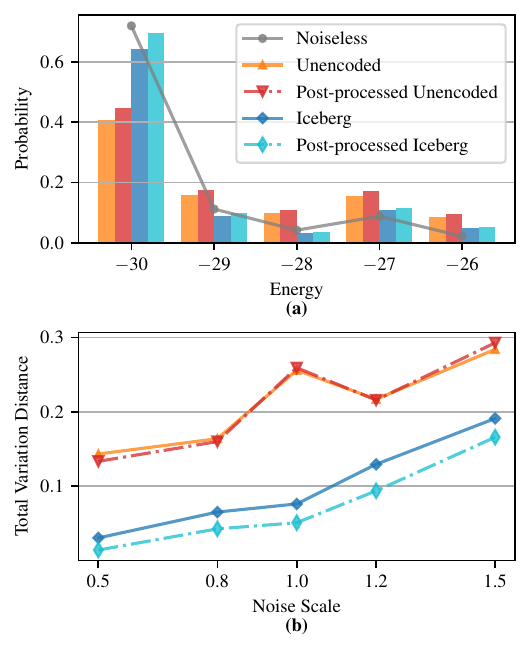}
    \caption{
    Iceberg enhances the benchmarking of QAOA. 
    (a) The energy probabilities associated with the $p=10$ QAOA states of a $k=22$ instance. The proposed Iceberg-QAOA captures the energy distribution more accurately than the unencoded QAOA execution and thus achieves the algorithmic break-even.
    (b) TVD analysis with different noise levels. As the noise level decreases, the Iceberg encoding captures the energy probability more accurately. Post-processing is more beneficial for Iceberg QAOA, with relative improvements ranging from $13\%$ to $53\%$, compared to $-3\%$ to $7\%$ for unencoded QAOA.
    }
    \label{fig:energy_population}
\end{figure}
\begin{figure}[t]
    \centering
    \includegraphics[width=1\linewidth]{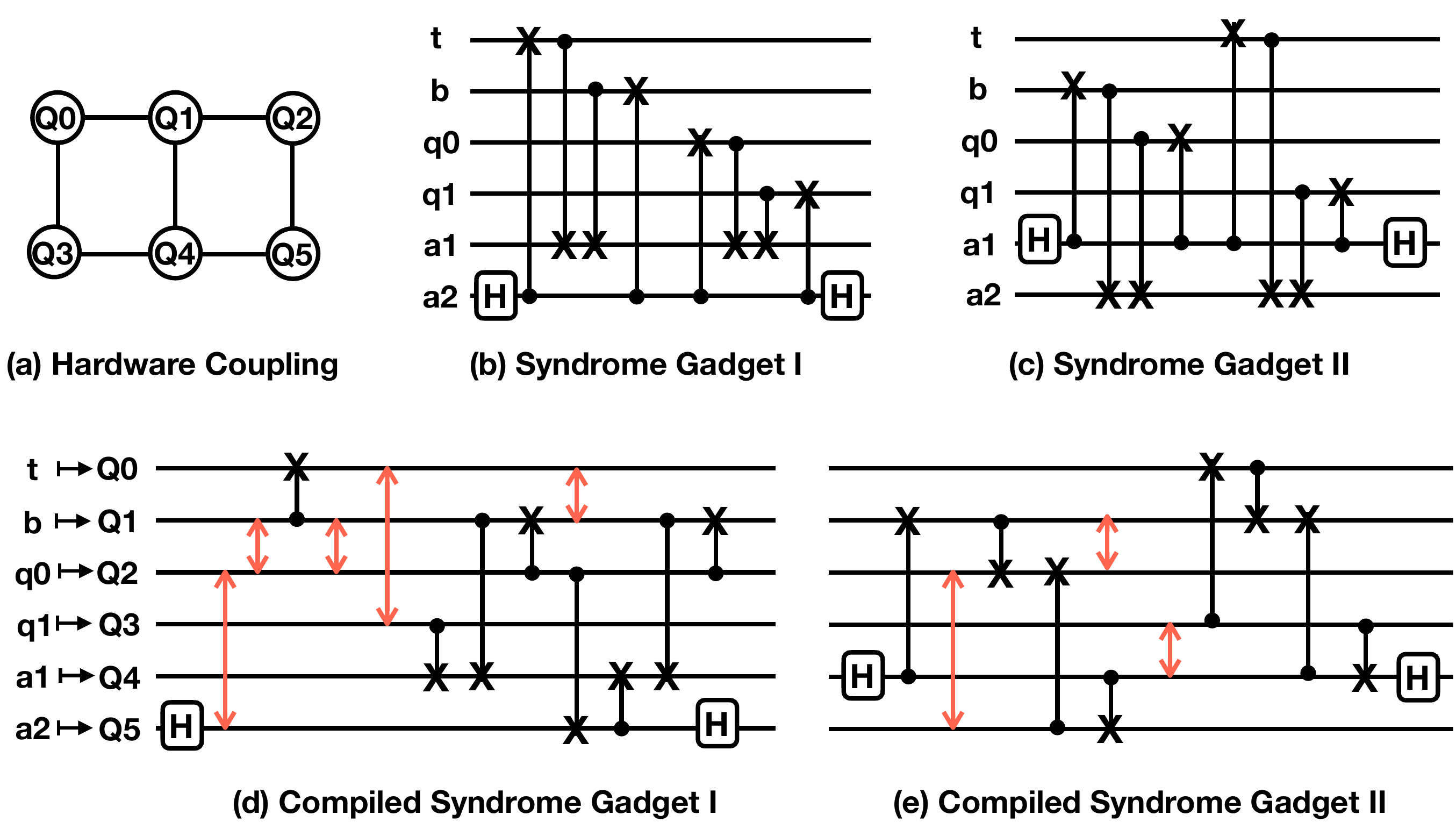}
    \caption{ \update{Syndrome gadget compilation with hardware connectivity constraints. (a) The hardware coupling graph with limited connectivity. (b) The smallest syndrome gadget following the original construction. (c) A hardware-friendly syndrome gadget with the same number of CNOT gates. (d) With the given initial mapping, the compiled syndrome gadget~\RNum{1} costs 5 SWAP gates. (e) With the same initial mapping, the compiled syndrome gadget~\RNum{2} requires only 3 SWAP gates.
     } }
    \label{fig:flex_hardware}
\end{figure}

\section{Discussion}
QED is a powerful technique for extracting signals from noisy quantum circuits. \update{In the near term, QED can credibly be expected to support beyond-classical algorithmic experimentation on hardware. Here, we discuss an example of how the Iceberg code supports algorithm study, albeit at a modest and classically tractable scale. 

Inspired by the long-tail behavior shown in \Cref{fig:hardware_results_panel}~(c), we propose a simple post-processing strategy to obtain more accurate energy probabilities. Since we are typically interested in optimal and near-optimal solutions, we can truncate the energy distribution and renormalize the probabilities of low-energy states, thereby uniformly increasing their likelihood. This strategy can be applied to any energy distributions obtained from QAOA. However, given Iceberg's longer tail, the post-processing will be particularly beneficial. 
In \Cref{fig:energy_population}(a), we visualize the energy probability of a $k=22$-node 3-regular graph instance executed on the H2-1 emulator. The Iceberg QAOA closely resembles the noiseless distribution compared to the unencoded version. In \Cref{fig:energy_population}(b), for the same instance and encoded circuit as in (a), we adjust the noise level settings of the emulator by uniformly scaling the error rate of all error models. We use the total variation as our distance metric, defined as $\mathrm{TVD} (P,Q)={\frac {1}{2}}\sum _{x}|P(x)-Q(x)|$ for probability distributions $P$ and $Q$. We then show the distance between noisy and noiseless QAOA energy distributions under different scenarios. As expected, the distance to the noiseless state decreases as noise levels decrease across all QAOA scenarios. The post-processing achieves up to a 57\% improvement for the encoded circuit in TVD, compared to 12\% for the unencoded circuit.}

Although QED has demonstrated improvements in various scenarios, its effectiveness is often constrained by diminishing post-selection rates and persistent logical error rates. Recent evidence~\cite{he2024performance} indicates that mitigating memory errors can effectively enhance the performance of QED codes and broaden their range of applicability.

This paper introduces a co-compilation technique for QED and algorithmic circuits, significantly reducing the depth of encoded circuits through the design of novel fault-tolerant QED gadgets and by leveraging their inherent flexibility. 
The co-optimized Iceberg-encoded QAOA circuit is demonstrated on \update{the Quantinuum H2-1 trapped-ion quantum processor}, achieving superior results compared to the previous state-of-the-art~\cite{he2024performance}. Notably, the break-even point --- where the encoded circuit surpasses the unencoded one --- has been improved from \emph{20 qubits to 34 qubits}. 
In addition to algorithmic performance, an application of QED for characterizing the energy population of a QAOA state is presented, underscoring the potential of QED for benchmarking quantum algorithms on hardware.

In the long term, our advancements in compiling to the Iceberg code will remain valuable due to its wide usage in code concatenation, such as for lowering the qubit overhead of the surface code~\cite{gidney2025yoked}.
Furthermore, the co-compilation framework is broadly applicable to other codes and architectures.
Similar to existing generic compilers targeting nearest-neighbor architectures~\cite{gushu:qubitmapping,alwin:qubitMapping, bochen:qubitmapping, qubitMappingAstar} and domain-specific compilers that leverage high-level abstractions~\cite{qaoaMappingAstar, linlin:2qan, gushu:paulihedral, liu:quclear, jin:tetris} for circuit optimization, our framework can be extended to compile circuits for nearest-neighbor architectures with alternative objectives, such as reducing the SWAP gate count. \update{
As demonstrated in \Cref{fig:flex_hardware}, we compiled the smallest syndrome gadget with respect to a given sparsely connected hardware topology and initial qubit mapping. The original syndrome gadget compilation requires at least five SWAP gates, whereas the hardware-friendly syndrome gadget requires only three SWAP gates. This example illustrates the potential for extending our proposed compiler into other architectures beyond the all-to-all connected ones, which we reserve for future work.}

\bibliographystyle{plain}
\bibliography{Bib/reference.bib}

\section*{Disclaimer}
This paper was prepared for informational purposes by the Global Technology Applied Research center of JPMorgan Chase \& Co. This paper is not a product of the Research Department of JPMorgan Chase \& Co. or its affiliates. Neither JPMorgan Chase \& Co. nor any of its affiliates makes any explicit or implied representation or warranty and none of them accept any liability in connection with this paper, including, without limitation, with respect to the completeness, accuracy, or reliability of the information contained herein and the potential legal, compliance, tax, or accounting effects thereof. This document is not intended as investment research or investment advice, or as a recommendation, offer, or solicitation for the purchase or sale of any security, financial instrument, financial product or service, or to be used in any way for evaluating the merits of participating in any transaction.

\end{document}